\documentclass[aps,reprint,superscriptaddress,pra]{revtex4-2}
\usepackage[utf8]{inputenc}
\usepackage{amsmath}
\usepackage{hyperref}
\hypersetup{colorlinks=true, citecolor=blue}
\usepackage{tikz}
\usepackage{subfigure}
\usetikzlibrary{quantikz}

\begin{document}
\title{Gravitational-wave matched filtering on a quantum computer}

\author{Do\u{g}a Veske}
\email{veske@thphys.uni-heidelberg.de}
\affiliation{Department of Physics, Columbia University in the City of New York, New York, NY 10027, USA}
\affiliation{Institute for Theoretical Physics, Heidelberg University, Heidelberg 69120, Germany}
\author{Cenk T\"uys\"uz}
\affiliation{Deutsches Elektronen-Synchrotron DESY, Platanenallee 6, 15738,
Zeuthen, Germany}
\affiliation{Instit\"ut f\"ur Physik, Humboldt-Universit\"at zu Berlin,
Newtonstr. 15, 12489, Berlin, Germany}
\author{Mirko Amico}
\affiliation{IBM Quantum, IBM T.J. Watson Research Center, Yorktown Heights, NY, USA}
\author{Nicholas T. Bronn}
\affiliation{IBM Quantum, IBM T.J. Watson Research Center, Yorktown Heights, NY, USA}
\author{Olivia T. Lanes}
\affiliation{IBM Quantum, IBM T.J. Watson Research Center, Yorktown Heights, NY, USA}
\author{Imre Bartos}
\affiliation{Department of Physics, University of Florida, PO Box 118440, Gainesville, FL 32611-8440, USA}
\author{Zsuzsa M\'arka}
\affiliation{Columbia Astrophysics Laboratory, Columbia University in the City of New York, New York, NY 10027, USA}
\author{Sebastian Will}
\affiliation{Department of Physics, Columbia University in the City of New York, New York, NY 10027, USA}
\author{Szabolcs M\'arka}
\affiliation{Department of Physics, Columbia University in the City of New York, New York, NY 10027, USA}

\begin{abstract}
State of the art quantum computers have very limited applicability for accurate calculations. Here we report the first experimental demonstration of qubit-based matched filtering for a detection of the gravitational-wave signal from a binary black hole merger. With our implementation on noisy superconducting qubits, we obtained a similar signal-to-noise ratio for the binary black hole merger as achievable with classical computation, providing evidence for the utility of qubits for practically relevant tasks. The algorithm we invented for this application is a Monte Carlo algorithm which uses quantum and classical computation together. It provides a quasi-quadartic speed-up for time-domain convolution, similar to achievable with fast Fourier transform.
\end{abstract}
\maketitle
\section{Introduction}
Recognizing patterns in large, noisy data sets is a key computational challenge with broad applications such as in signal detection, image analysis and machine learning. With the increasing amount of collected data, improvements on the execution of such analyses will improve our capability to process it. Such improvements may come in many forms from faster calculation algorithms to employing more energy efficient hardware.

Knowing the precise shape of the signal one searches for in a dataset increases the statistical power of the search compared to a blind search for an anomaly. The optimal search technique for finding a known signal buried in Gaussian noise is called matched filtering~\cite{leon}. Originally developed for detecting radar echoes in the 1950's~\cite{WOODWARD195381}, its approximate optimality in many real world applications made it the method of choice in many different areas from gravitational-wave (GW) detection~\cite{Allen_2012,PhysRevLett.116.061102} to quantum tomography via correlation of noisy measurements~\cite{Ryan2015}. 

For a known digitized template signal $x_i$ with $N$ points and a data stream $y_i$ with $L$ points ($N\leq L$) in the presence of white noise, in time domain, the signal-to-noise ratio (SNR) of a matched filter can be computed with the convolution
\begin{equation}
    \rho[j]=\sum_{i=1}^N y_{i+j}x_i
    \label{eq:mf2}
\end{equation}
In the presence of colored noise, $x$ and $y$ additionally need to be normalized using the noise's power spectral density.

Calculations of the form of Eq. \eqref{eq:mf2} have time-complexity of $\mathcal{O}(NL)$ if done in time domain. It can be calculated more efficiently in frequency domain using fast Fourier transform (FFT) with $\mathcal{O}(L\log N)$ time complexity. Here we do this calculation with a similar efficiency without using FFT and with a non-conventional hardware, a quantum computer. We developed a hybrid algorithm that uses classical and quantum computers together to calculate the sums of the form Eq. \eqref{eq:mf2} quasi-quadratically faster than a classical computer for real valued functions; its time complexity scales as $\mathcal{O}(L(\log N)^2)$. This speed-up in our algorithm is not related to any mathematical symmetry of complex exponentials, which are the basis for FFT. The improvement in our algorithm comes from the utilization of quantum operations. With this algorithm, we performed the first qubit-based matched filtering with a quantum computer. Currently there is no clear advantage of using this hybrid algorithm over using FFT as they both have similar time complexities. However, it is hard to predict the future advantages that may come from the use of a conceptually different hardware and algorithm. As a speculation, sharing the computational load to classical and quantum computers may be more energy efficient in the future.

The algorithm presented here is a Monte Carlo approach to solving the problem. We leverage the divide-and-conquer algorithm~\cite{Araujo_2021} to encode a real valued vector on the quantum computer. The only quantum operations we perform in our method are encoding of the problem into a string of qubits and performing quantum measurements on them. These operations, without additional quantum computation can, by themselves, provide a quantum advantage. Quantum encoding and measurement can generate random bitstrings according to a probability distribution exponentially faster than classical generation by measuring the end states of amplitude encoding. We basically exploit this attribute. The divide-and-conquer algorithm~\cite{Araujo_2021} requires poly-log depth circuits and since we don't perform any other operation, this use of low-depth quantum circuits makes our method more robust against the noise of the quantum hardware without any error correction. This is the main reason why we were able to demonstrate our method on real quantum hardware with high accuracy. 

Our experimental implementation we present in this letter, provide another evidence for the use of quantum computers on practically useful tasks, which are not specifically crafted for quantum computers. We executed our experiment on a GW signal from a binary black hole merger. Template based GW search algorithms are searches employing matched filtering using millions of templates~\cite{cannon2020gstlal,Nitz_2021}, where known waveforms of merging binary systems of black holes or neutron stars, which are computed via Einstein's field equations, are used~\cite{Allen_2012,PhysRevLett.116.061102}. By the end of the third observing run of the LIGO~\cite{aligo2015} and Virgo~\cite{Acernese_2014} detectors, about $90$ astrophysical gravitational-wave (GW) discoveries have been reported~\cite{gwtc1,gwtc2,gwtc21,gwtc3}. Besides the detectors' unprecedented sensitivity, observations succeeded thanks to the deployed state-of-the-art search techniques. This computational load in GW astronomy, has attracted help from quantum computation too; search and analysis algorithms have been developed which use quantum computation~\cite{miyamoto2022gravitational,PhysRevResearch.4.023006,Escrig_2023}. 

Below, we first briefly explain the basic principles of our method. Then, we explain our implementation and show our results.

\section{Method} First, the signal template and data are encoded to independent sets of qubits via amplitude encoding which can encode $n$ complex numbers to the amplitudes of the $n$ different computational basis states spanned by $\log_2n$ qubits. Despite the logarithmic gain in the number of qubits, the computational complexity of amplitude encoding scales with $\mathcal{O}(n)$ which at first sight seems to negate any gain. However, it has been shown that with the use of extra qubits (typically called ancilla or auxiliary qubits) during the encoding stage, which are thrown after the encoding is done, the time complexity of amplitude encoding can be reduced~\cite{https://doi.org/10.48550/arxiv.2201.11495}. Specifically Ref.~\cite{Araujo_2021} discusses a divide and conquer algorithm which uses $\mathcal{O}(n)$ qubits to encode $\log_2n$ qubits with a circuit depth $\mathcal{O}((\log n)^2)$. This improvement enables advantageous uses of qubits for computation without losing the advantage in the beginning. We use this encoding algorithm in our method. In Appendix \ref{sec:appa} we briefly explain how this algorithm works.

We encode the data and signal such that the coefficients of the computational basis states of the qubit states are proportional to the square root of the corresponding values of the signal and data, which have been offset by the absolute value of their minimum negative values to make them all positive. The overall encoded state $\ket{\psi}=\ket{x}\ket{y}$ is formed by
\begin{subequations}
\begin{equation}
    {\rm Data}=\ket{y}=\mathcal{N}_y^{-1}\sum_{i=1}^{L} \sqrt{y_i+\Delta y}\ket{i}_y,
\end{equation}
\begin{equation}
    {\rm Signal}=\ket{x}=\mathcal{N}_x^{-1}\sum_{i=1}^{N} \sqrt{x_i+\Delta x}\ket{i}_x
\end{equation}
\end{subequations}
where $\ket{i}$ are the $2^n$ possible basis states obtainable with $n$ qubits, i.e. $\ket{00}$, $\ket{01}$, $\ket{10}$, $\ket{11}$ for 2 qubits; $\Delta x=-\min(x_i)$ and $\Delta y=-\min(y_i)$, with normalizations $\mathcal{N}_x=\sum_{i=1}^N (x_i+\Delta x)$ and $\mathcal{N}_y=\sum_{i=1}^L (y_i+\Delta y)$.

In the next step, joint measurements of the encoded independent sets of data and signal qubits generate bitstrings whose appearance probabilities are proportional to the product of the offset signal and data values for the times (indices) each bitstring uniquely represent. After the measurement, a classical logic circuit relocates each bitstring to their correct place (index) in the output, according to Eq.~\eqref{eq:mf2}, and stacks every occurrence. Averaging over series of operations of the circuit samples possible occurrences and constructs the desired discrete output signal. Due to the offset of data and signal template, and the normalizations a correction is done at the very end. These operations correspond to calculating the SNR as

\begin{multline}
    \rho[j]=\mathcal{N}_y\mathcal{N}_x\sum_{i=1}^N |\bra{i+j}_y\bra{i}_x\ket{\psi}|^2\\-\sum_{i=1}^N\Delta yx_i+\Delta xy_{i+j}+\Delta y\Delta x
    \label{eq:corrected}
\end{multline}

\begin{figure*}
    \centering
    \includegraphics[width=\textwidth]{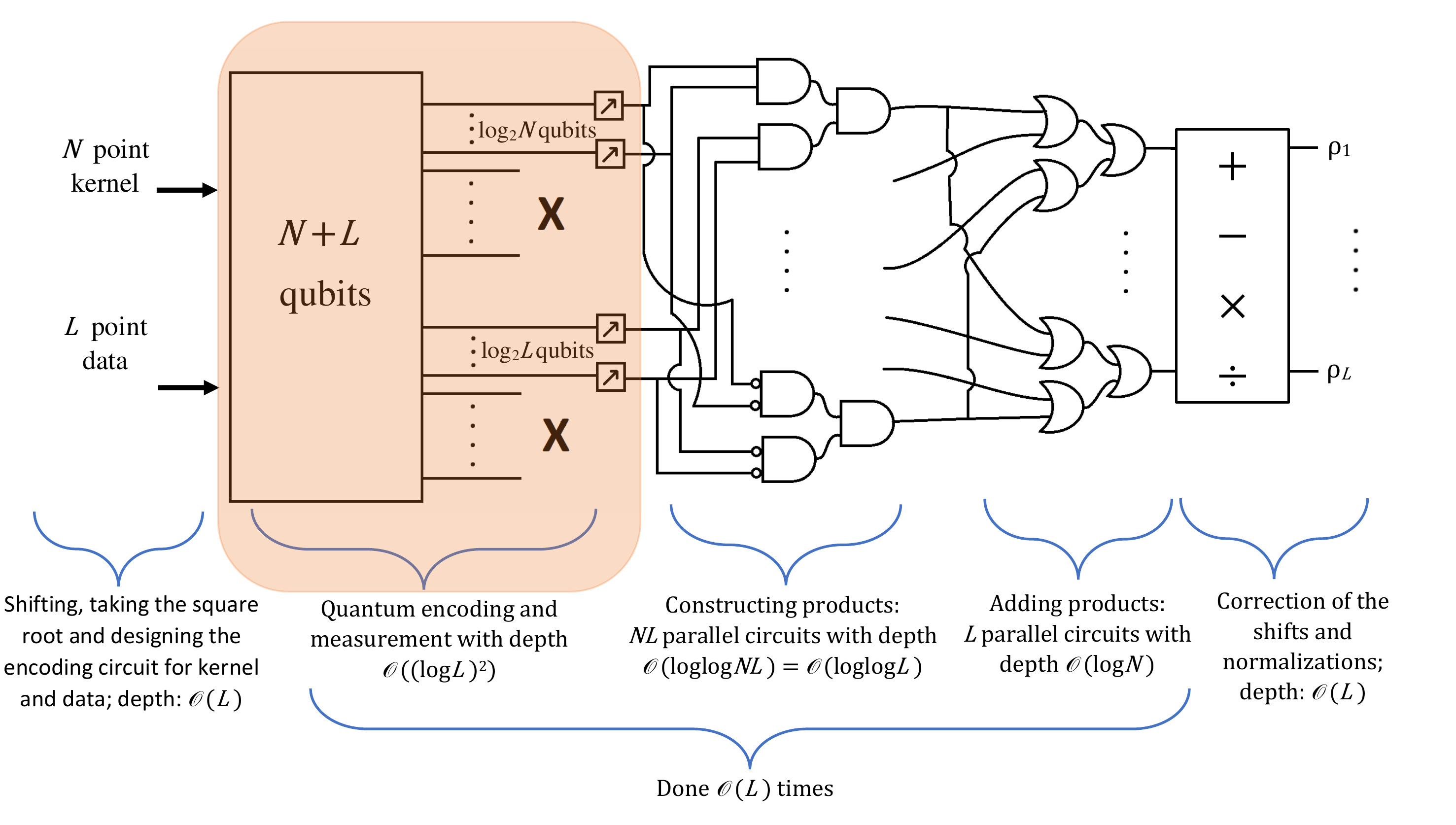}
    \caption{The convolution gate describing the hybrid method and showing the asymptotic computation times of its parts, without optimizing by dividing the data into smaller segments. The shaded region points out the part with quantum operations. The total time complexity which is $\mathcal{O}(L(\log L)^2)$ in the figure reduces to $\mathcal{O}(L(\log N)^2)$ after that optimization (see Appendix \ref{sec:appb}).}
    \label{fig:qcfig}
\end{figure*}

Among the time complexities of each of these classical logical, mathematical and quantum operations; the dominant one is of the amplitude encoding which is $\mathcal{O}((\log L)^2)$ with the divide and conquer algorithm~\cite{Araujo_2021}. Moreover, in order to have a fixed precision of each SNR value, one needs to have certain amount of shots per calculated SNR. This necessitates the total shot count to scale with the number of SNRs to be calculated, i.e. $\mathcal{O}(L)$. Therefore, the overall time complexity of this algorithm is $\mathcal{O}(L(\log L)^2)$. Due to worse scaling than linear scaling, an additional optimization can be done by dividing the data into smaller segments which can reduce the overall time complexity to $\mathcal{O}(L(\log N)^2)$. The precision of each SNR estimate is

\begin{multline}
    \frac{\delta \rho[j]}{\rho[j]}=\sqrt{\frac{\sum_{i\neq j}\rho[i]}{\rho[j]\times s\times L}}\\ \times\left(1+\frac{\sum_{i=1}^N\Delta yx_i+\Delta xy_{i+j}+\Delta y\Delta x}{\rho[j]}\right)
\end{multline}
where $s$ is the total shot count divided by $L$. More details on precision, time complexity of each operation, and data length manipulation for extra optimization are explained in the Appendix \ref{sec:appb}. The whole algorithm can be implemented on quantum and classical hardware and can be seen as a convolution gate. The complete block diagram of the circuit is shown in Fig.~\ref{fig:qcfig}. A similar hybrid construction of the SWAP test~\cite{PhysRevLett.87.167902,kang2019implementation}, which computes the inner product of two vectors, with the purpose of simplifying the original quantum circuit with classical additions exists in Ref.~\cite{Cincio_2018}, which we were inspired from. We also note that if shallower amplitude encoding algorithms are invented in the future, which were theoretically proven to exist~\cite{https://doi.org/10.48550/arxiv.2201.11495}, our method's total time complexity can be further reduced down to $\mathcal{O}(L(\log N))$.

\section{Implementation}
We experimentally demonstrated the operation of our method by accessing IBM Quantum systems through the cloud via Qiskit~\cite{treinish_qiskit_2022}. We computed the SNR time series of the astrophysical gravitational-wave event GW190521 ~\cite{PhysRevLett.125.101102} for a binary black hole merger gravitational waveform and demonstrated the accuracy of the method by comparing this result with the output of classical computation (see also Appendix \ref{sec:appc} for a preliminary study on artificial random data). IBM Quantum backends consist of fixed-frequency superconducting transmons~\cite{Koch2007} coupled by transmission line ``bus'' resonators arranged in a heavy-hexagonal lattice~\cite{Chamberland2019}. Single-qubit gates are achieved by on-resonant microwave drives at the ground-to-excited transition frequencies of the transmons, which vary due to the amorphous oxide of the single Josephson junction that forms the inductive part of the transmon (large superconducting pads form the geometric shunting capacitor of the analogous anharmonic oscillator). The cross-resonance entangling gate~\cite{Chow2011, Sheldon2016} uses these variations in frequency to create a microwave-activated drive that entangles coupled transmons and forms the basis of the CNOT gate. Measurement is done in the dispersive regime of circuit quantum electrodynamics~\cite{Blais2004, Wallraff2004}, in which a readout resonator is populated by photons far detuned from the qubit frequency, which dephase each qubit and imparts a state-dependent phase shift on the measurement pulse. The available native basis gates on these backends are highly-calibrated $\pi/2$ rotations \texttt{sx} gates, virtual-Z~\cite{McKay2017} \texttt{rz} gates for single qubit operations; and CNOT \texttt{cx} gates for multi-qubit operations. The other multi-qubit operations are executed with a combination of these gates. For an extended discussion on the architecture of the systems with superconducting qubits we refer to the Refs. \cite{Gambetta_2017,Bravyi_2022}. Sources of error in these operations are available~\cite{IQS} and show the main sources of error to be the CNOT error (average of 1\%) and readout assignment error (1-4\%). As deeper quantum circuits tend to have more CNOTs, these tend to be the biggest source of error. Due to limited physical connectivity, SWAP gates (consisting of 3 CNOTs) are particularly costly when mapping algorithms to current noisy quantum hardware.

Due to these limitations, we non-optimally divided the GW data ($y$) into segments of length $k_d=4$  and the template ($x$) into segments of length $k_t=2$. The main obstacle for not using more data points at once is the significantly greater estimated error in the encoding when data is encoded to more than 2 qubits with the divide and conquer algorithm~\cite{Araujo_2021}, due to the need of controlled SWAP operations between non-connected qubits. The template lengths (=2) were chosen less than the data lengths (=4) in order to have more than one SNR value calculated with the described method. In this case there are 3 SNR values calculated per one data set. They correspond to the measurement probabilities of the template ($x$) and data ($y$) qubit probabilities $P(\ket{0}_x\ket{00}_y)+P(\ket{1}_x\ket{01}_y)$, $P(\ket{0}_x\ket{01}_y)+P(\ket{1}_x\ket{10}_y)$ and $P(\ket{0}_x\ket{10}_y)+P(\ket{1}_x\ket{11}_y)$.

\begin{figure*}
\centering
\subfigure[]{\includegraphics[width=0.49\textwidth]{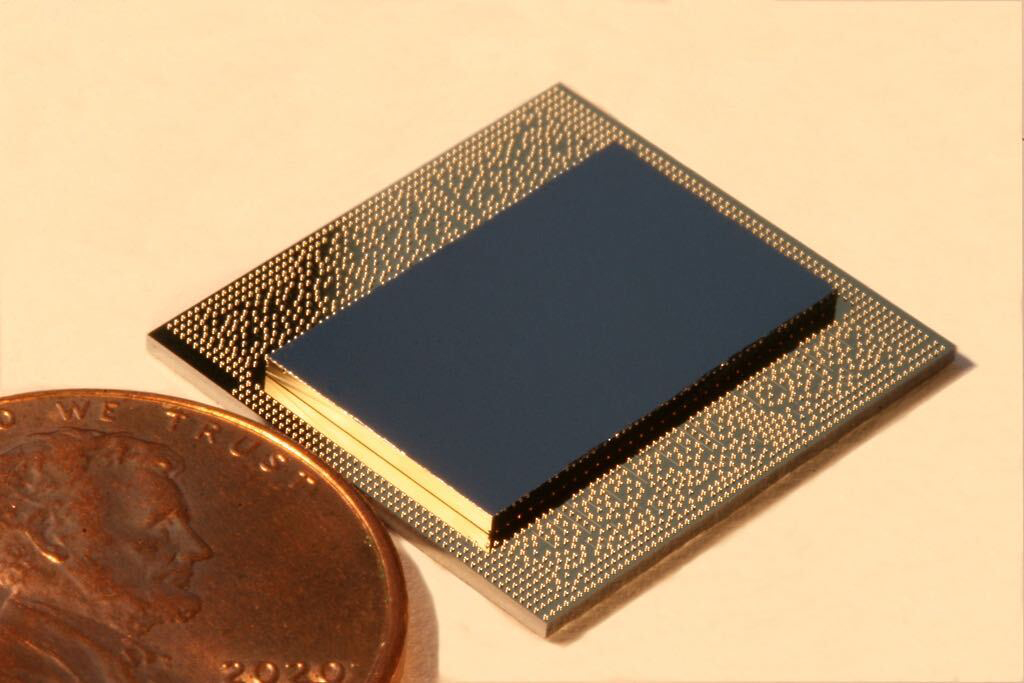}}
\subfigure[]{\includegraphics[width=0.49\textwidth]{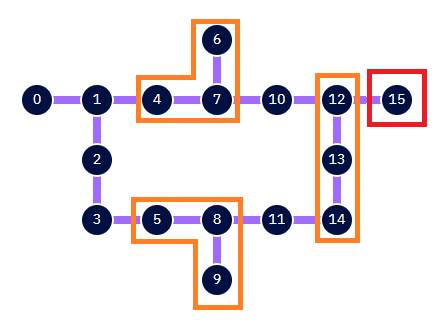}}
    \caption{(a) Photo of the Falcon chip, the model of $ibmq\_guadalupe$ , next to a US penny (credit: IBM Research); (b)
    the qubit layout of the $ibmq\_guadalupe$. The qubits we used for encoding the data are encapsulated with orange lines and the qubit we used for encoding the signal template is encapsulated with red lines. These qubits were chosen due to the lowest CNOT and readout error rates at the time of execution.}
    \label{fig:guadalupe}
\end{figure*}

For our demonstration with the real gravitational-wave data, we have specifically chosen the event GW190521~\cite{PhysRevLett.125.101102}. It consists of relatively low gravitational-wave frequencies compared to the observed duration of the signal as it is one of the heaviest stellar-mass binary black hole systems ever detected by LIGO Scientific Collaboration and Virgo Collaboration~\cite{gwtc1,gwtc2,gwtc21,gwtc3}. This allows us to down-sample the detected data series to reduce the data points to be processed with the quantum hardware. We used the 32 s long aLIGO Livingston~\cite{TheLIGOScientific:2014jea} detector's data sampled at 4 kHz \footnote{\url{https://www.gw-openscience.org/eventapi/html/GWTC-2/GW190521/}}. We down-sampled it to 200 Hz, after digitally applying an ideal low-pass filter with the cutoff frequency 99.98 Hz which is above the maximum frequency in the reconstructed astrophysical gravitational-wave. The noise power spectral density was found from the original data with Welch's average periodogram method~\cite{1161901} with segment lengths of 512, using the \texttt{matplotlib}~\cite{4160265} python package's \texttt{matplotlib.pyplot.psd} function. According to the estimation of the source properties of the gravitational-wave event at the detector frame~\cite{Abbott_2020}; the template waveform was generated at 200 Hz sampling with the \texttt{gwsurrogate}~\cite{PhysRevX.4.031006} package with the NRSur7dq4 waveforms~\cite{2019PhRvR...1c3015V}. The used inputs were 154.7 solar masses and 120.1 solar masses for the detector frame black hole masses, (0.69,0,0) and (0,0.73,0) for the dimensionless spin vectors, and 12 Hz for the start frequency. With these configurations the SNRs were calculated for the data points starting at the UTC time 1242442967.15 for the consequent 0.45 s. The experiments were ran on the backend {\it ibmq\_guadalupe}~\cite{IQS}. In each run, from the 16 qubits on the backend, up to 10 qubits were used for independently encoding 3 partial data sets with 3 qubits each and a template part with 1 qubit in parallel. In this configuration, total of 308 circuits were ran for covering the full signal, each with $10^4$ shots. The photo of the Falcon chip, which is the model of {\it ibmq\_guadalupe}, and the layout of {\it ibmq\_guadalupe} with the used qubits are shown in Fig.~\ref{fig:guadalupe}. At the time of execution, the CNOT error probabilities of each qubit connection and readout error probabilities of the qubits were all about 1\%. The SNR for a particular time was found by adding the SNRs of the 2 point partial templates, which overlap with the previous and next partial template at 1 point. 

\section{Results}
\begin{figure}
    \centering
    \includegraphics[width=\columnwidth]{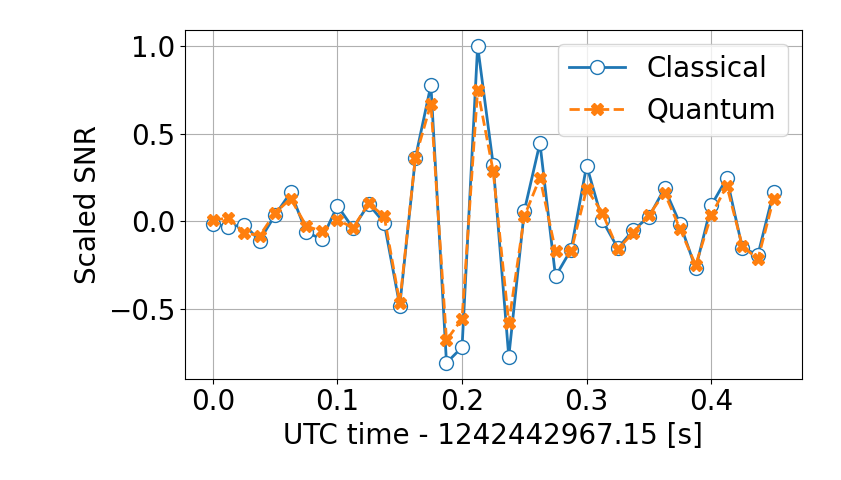}
    \caption{Scaled SNR time series around the gravitational-wave event from black hole merger GW190521. Blue circles show the classical result and orange crosses show the results obtained with the hybrid algorithm.}
    \label{fig:gw}
\end{figure}

The 308 quantum circuits each with $10^4$ shots took 14.5 minutes in total to run. The SNR series obtained with the quantum processor along with the pure classical computation are shown in Fig.~\ref{fig:gw}. For an easier \emph{relative} comparison of the quantum processor's performance, they were scaled in the figure such that the classically computed maximum SNR reaches the value of 1; since the overall factors are calculated classically in the hybrid calculation and do not change the relative accuracy. The hybrid calculation shows high accuracy. The main deviation between the two methods is the slight reduction of highest absolute SNR values, which shows the effect of the non-idealities of the current hardware (see Appendix \ref{sec:appc} for our similar findings with artificial data). This shows that current quantum hardware can be used for performing calculations with comparable accuracy to a classical computer for practically relevant problems. 

With technological improvements, these calculations can be performed faster with quantum computers. The reasons why quantum advantage could not be achieved at this point are the CNOT error rates and the limited connectivity between the qubits which increases the required number of CNOT gates for encoding. These prohibit the construction of more intricate configurations for encoding larger number of qubits, which would have advantage over classical computation. Executing circuits with large number of qubits remains to be a hardware challenge in the noisy intermediate-scale quantum era~\cite{kechedzhi2023effective}. With future hardware improvements, a useful application of this method can be executed.

\section{Outlook}
We point out that the described method for matched filtering is straightforward to generalize for any calculations involving multi-dimensional functions; e.g., in image processing, while obeying the same scaling, where $N$ and $L$ represent the total number of points in the higher dimensional spaces, since any higher dimensional finite number of samples can be represented in one dimensional series. A corresponding one dimensional filter can be found according to the rasterizing scheme to one dimension. Moreover, our method can be generalized to time independent system responses or integral transforms having more than one input or more than one instances of the same input as in non-linear systems, Volterra series or coherent matched filtering of multi-detector data. These constructions with more than one input are simply special cases of multi-dimensional functions.

Our method uses quantum and classical computation complementarily instead of relying completely on quantum computation. This makes it more resilient to the typical gate errors due to having shallow quantum circuits and hence makes it a method that can be used reliably in the nearest future compared to the methods that require deeper quantum circuits. The presented demonstration proved the method's experimental viability on IBM Quantum's hardware and which can have an advantage with improving technology, especially with fully connected qubit networks. Although performing similarly to the classical computation with FFT, in the future distributing the computing load to quantum and classical computers may have other practical benefits such as energy saving or hardware scalability.

\section*{Acknowledgements}
This document was reviewed by LIGO Scientific Collaboration under the document number P2200255. Authors are grateful to Mesut Çalışkan for this review, detailed inspection and helpful comments. Authors thank Columbia University in the City of New York for its unwavering and generous support. D.V. thanks Baran Bodur for suggesting the name convolution gate. D.V. acknowledges the support by European Research Council (ERC) under the European Union's Horizon 2020 research and innovation programme grant agreement No 801781. C.T. is supported in part by the Helmholtz Association - “Innopool Project Variational Quantum Computer Simulations (VQCS)”. I.B. acknowledges the support of the Alfred P. Sloan Foundation. We acknowledge the use of IBM Quantum services for this work. The views expressed are those of the authors, and do not reflect the official policy or position of IBM or the IBM Quantum team. This research has made use of data or software obtained from the Gravitational Wave Open Science Center (gwosc.org)~\cite{RICHABBOTT2021100658,Abbott_2023}, a service of the LIGO Scientific Collaboration, the Virgo Collaboration, and KAGRA. This material is based upon work supported by NSF's LIGO Laboratory which is a major facility fully funded by the National Science Foundation, as well as the Science and Technology Facilities Council (STFC) of the United Kingdom, the Max-Planck-Society (MPS), and the State of Niedersachsen/Germany for support of the construction of Advanced LIGO and construction and operation of the GEO600 detector. Additional support for Advanced LIGO was provided by the Australian Research Council. Virgo is funded, through the European Gravitational Observatory (EGO), by the French Centre National de Recherche Scientifique (CNRS), the Italian Istituto Nazionale di Fisica Nucleare (INFN) and the Dutch Nikhef, with contributions by institutions from Belgium, Germany, Greece, Hungary, Ireland, Japan, Monaco, Poland, Portugal, Spain. KAGRA is supported by Ministry of Education, Culture, Sports, Science and Technology (MEXT), Japan Society for the Promotion of Science (JSPS) in Japan; National Research Foundation (NRF) and Ministry of Science and ICT (MSIT) in Korea; Academia Sinica (AS) and National Science and Technology Council (NSTC) in Taiwan.

\bibliographystyle{apsrev4-2}
\bibliography{refs}

\begin{thebibliography}{44}%
\makeatletter
\providecommand \@ifxundefined [1]{%
 \@ifx{#1\undefined}
}%
\providecommand \@ifnum [1]{%
 \ifnum #1\expandafter \@firstoftwo
 \else \expandafter \@secondoftwo
 \fi
}%
\providecommand \@ifx [1]{%
 \ifx #1\expandafter \@firstoftwo
 \else \expandafter \@secondoftwo
 \fi
}%
\providecommand \natexlab [1]{#1}%
\providecommand \enquote  [1]{``#1''}%
\providecommand \bibnamefont  [1]{#1}%
\providecommand \bibfnamefont [1]{#1}%
\providecommand \citenamefont [1]{#1}%
\providecommand \href@noop [0]{\@secondoftwo}%
\providecommand \href [0]{\begingroup \@sanitize@url \@href}%
\providecommand \@href[1]{\@@startlink{#1}\@@href}%
\providecommand \@@href[1]{\endgroup#1\@@endlink}%
\providecommand \@sanitize@url [0]{\catcode `\\12\catcode `\$12\catcode
  `\&12\catcode `\#12\catcode `\^12\catcode `\_12\catcode `\%12\relax}%
\providecommand \@@startlink[1]{}%
\providecommand \@@endlink[0]{}%
\providecommand \url  [0]{\begingroup\@sanitize@url \@url }%
\providecommand \@url [1]{\endgroup\@href {#1}{\urlprefix }}%
\providecommand \urlprefix  [0]{URL }%
\providecommand \Eprint [0]{\href }%
\providecommand \doibase [0]{https://doi.org/}%
\providecommand \selectlanguage [0]{\@gobble}%
\providecommand \bibinfo  [0]{\@secondoftwo}%
\providecommand \bibfield  [0]{\@secondoftwo}%
\providecommand \translation [1]{[#1]}%
\providecommand \BibitemOpen [0]{}%
\providecommand \bibitemStop [0]{}%
\providecommand \bibitemNoStop [0]{.\EOS\space}%
\providecommand \EOS [0]{\spacefactor3000\relax}%
\providecommand \BibitemShut  [1]{\csname bibitem#1\endcsname}%
\let\auto@bib@innerbib\@empty
\bibitem [{\citenamefont {Couch}(2012)}]{leon}%
  \BibitemOpen
  \bibfield  {author} {\bibinfo {author} {\bibfnamefont {L.~W.}\ \bibnamefont
  {Couch}},\ }\bibinfo {title} {Digital and analog communication systems}\
  (\bibinfo  {publisher} {Pearson},\ \bibinfo {year} {2012})\ Chap.\ \bibinfo
  {chapter} {6-8},\ \bibinfo {edition} {8th}\ ed.\BibitemShut {Stop}%
\bibitem [{\citenamefont {WOODWARD}(1953)}]{WOODWARD195381}%
  \BibitemOpen
  \bibfield  {author} {\bibinfo {author} {\bibfnamefont {P.}~\bibnamefont
  {WOODWARD}},\ }in\ \href
  {https://doi.org/https://doi.org/10.1016/B978-0-08-011006-6.50011-0} {\emph
  {\bibinfo {booktitle} {Probability and Information Theory with Applications
  to Radar (Second Edition)}}},\ \bibinfo {series and number} {International
  Series of Monographs on Electronics and Instrumentation},\ \bibinfo {editor}
  {edited by\ \bibinfo {editor} {\bibfnamefont {P.}~\bibnamefont {WOODWARD}}}\
  (\bibinfo  {publisher} {Pergamon},\ \bibinfo {year} {1953})\ \bibinfo
  {edition} {second edition}\ ed.,\ pp.\ \bibinfo {pages} {81--99}\BibitemShut
  {NoStop}%
\bibitem [{\citenamefont {Allen}\ \emph {et~al.}(2012)\citenamefont {Allen},
  \citenamefont {Anderson}, \citenamefont {Brady}, \citenamefont {Brown},\ and\
  \citenamefont {Creighton}}]{Allen_2012}%
  \BibitemOpen
  \bibfield  {author} {\bibinfo {author} {\bibfnamefont {B.}~\bibnamefont
  {Allen}}, \bibinfo {author} {\bibfnamefont {W.~G.}\ \bibnamefont {Anderson}},
  \bibinfo {author} {\bibfnamefont {P.~R.}\ \bibnamefont {Brady}}, \bibinfo
  {author} {\bibfnamefont {D.~A.}\ \bibnamefont {Brown}},\ and\ \bibinfo
  {author} {\bibfnamefont {J.~D.~E.}\ \bibnamefont {Creighton}},\ }\bibfield
  {journal} {\bibinfo  {journal} {Physical Review D}\ }\textbf {\bibinfo
  {volume} {85}},\ \href {https://doi.org/10.1103/physrevd.85.122006}
  {10.1103/physrevd.85.122006} (\bibinfo {year} {2012})\BibitemShut {NoStop}%
\bibitem [{\citenamefont {Abbott}\ \emph {et~al.}(2016)\citenamefont {Abbott},
  \citenamefont {Abbott}, \citenamefont {Abbott}, \citenamefont {Abernathy},
  \citenamefont {Acernese}, \citenamefont {Ackley}, \citenamefont {Adams},
  \citenamefont {Adams}, \citenamefont {Addesso}, \citenamefont {Adhikari}
  \emph {et~al.}}]{PhysRevLett.116.061102}%
  \BibitemOpen
  \bibfield  {author} {\bibinfo {author} {\bibfnamefont {B.~P.}\ \bibnamefont
  {Abbott}}, \bibinfo {author} {\bibfnamefont {R.}~\bibnamefont {Abbott}},
  \bibinfo {author} {\bibfnamefont {T.~D.}\ \bibnamefont {Abbott}}, \bibinfo
  {author} {\bibfnamefont {M.~R.}\ \bibnamefont {Abernathy}}, \bibinfo {author}
  {\bibfnamefont {F.}~\bibnamefont {Acernese}}, \bibinfo {author}
  {\bibfnamefont {K.}~\bibnamefont {Ackley}}, \bibinfo {author} {\bibfnamefont
  {C.}~\bibnamefont {Adams}}, \bibinfo {author} {\bibfnamefont
  {T.}~\bibnamefont {Adams}}, \bibinfo {author} {\bibfnamefont
  {P.}~\bibnamefont {Addesso}}, \bibinfo {author} {\bibfnamefont {R.~X.}\
  \bibnamefont {Adhikari}}, \emph {et~al.} (\bibinfo {collaboration} {LIGO
  Scientific Collaboration and Virgo Collaboration}),\ }\href
  {https://doi.org/10.1103/PhysRevLett.116.061102} {\bibfield  {journal}
  {\bibinfo  {journal} {Phys. Rev. Lett.}\ }\textbf {\bibinfo {volume} {116}},\
  \bibinfo {pages} {061102} (\bibinfo {year} {2016})}\BibitemShut {NoStop}%
\bibitem [{\citenamefont {Ryan}\ \emph {et~al.}(2015)\citenamefont {Ryan},
  \citenamefont {Johnson}, \citenamefont {Gambetta}, \citenamefont {Chow},
  \citenamefont {da~Silva}, \citenamefont {Dial},\ and\ \citenamefont
  {Ohki}}]{Ryan2015}%
  \BibitemOpen
  \bibfield  {author} {\bibinfo {author} {\bibfnamefont {C.~A.}\ \bibnamefont
  {Ryan}}, \bibinfo {author} {\bibfnamefont {B.~R.}\ \bibnamefont {Johnson}},
  \bibinfo {author} {\bibfnamefont {J.~M.}\ \bibnamefont {Gambetta}}, \bibinfo
  {author} {\bibfnamefont {J.~M.}\ \bibnamefont {Chow}}, \bibinfo {author}
  {\bibfnamefont {M.~P.}\ \bibnamefont {da~Silva}}, \bibinfo {author}
  {\bibfnamefont {O.~E.}\ \bibnamefont {Dial}},\ and\ \bibinfo {author}
  {\bibfnamefont {T.~A.}\ \bibnamefont {Ohki}},\ }\href
  {https://doi.org/10.1103/PhysRevA.91.022118} {\bibfield  {journal} {\bibinfo
  {journal} {Physical Review A}\ }\textbf {\bibinfo {volume} {91}},\ \bibinfo
  {pages} {022118} (\bibinfo {year} {2015})}\BibitemShut {NoStop}%
\bibitem [{\citenamefont {Araujo}\ \emph {et~al.}(2021)\citenamefont {Araujo},
  \citenamefont {Park}, \citenamefont {Petruccione},\ and\ \citenamefont
  {da~Silva}}]{Araujo_2021}%
  \BibitemOpen
  \bibfield  {author} {\bibinfo {author} {\bibfnamefont {I.~F.}\ \bibnamefont
  {Araujo}}, \bibinfo {author} {\bibfnamefont {D.~K.}\ \bibnamefont {Park}},
  \bibinfo {author} {\bibfnamefont {F.}~\bibnamefont {Petruccione}},\ and\
  \bibinfo {author} {\bibfnamefont {A.~J.}\ \bibnamefont {da~Silva}},\
  }\bibfield  {journal} {\bibinfo  {journal} {Scientific Reports}\ }\textbf
  {\bibinfo {volume} {11}},\ \href {https://doi.org/10.1038/s41598-021-85474-1}
  {10.1038/s41598-021-85474-1} (\bibinfo {year} {2021})\BibitemShut {NoStop}%
\bibitem [{\citenamefont {Cannon}\ \emph {et~al.}(2020)\citenamefont {Cannon},
  \citenamefont {Caudill}, \citenamefont {Chan}, \citenamefont {Cousins},
  \citenamefont {Creighton}, \citenamefont {Ewing}, \citenamefont {Fong},
  \citenamefont {Godwin}, \citenamefont {Hanna}, \citenamefont {Hooper},
  \citenamefont {Huxford}, \citenamefont {Magee}, \citenamefont {Meacher},
  \citenamefont {Messick}, \citenamefont {Morisaki}, \citenamefont {Mukherjee},
  \citenamefont {Ohta}, \citenamefont {Pace}, \citenamefont {Privitera},
  \citenamefont {de~Ruiter}, \citenamefont {Sachdev}, \citenamefont {Singer},
  \citenamefont {Singh}, \citenamefont {Tapia}, \citenamefont {Tsukada},
  \citenamefont {Tsuna}, \citenamefont {Tsutsui}, \citenamefont {Ueno},
  \citenamefont {Viets}, \citenamefont {Wade},\ and\ \citenamefont
  {Wade}}]{cannon2020gstlal}%
  \BibitemOpen
  \bibfield  {author} {\bibinfo {author} {\bibfnamefont {K.}~\bibnamefont
  {Cannon}}, \bibinfo {author} {\bibfnamefont {S.}~\bibnamefont {Caudill}},
  \bibinfo {author} {\bibfnamefont {C.}~\bibnamefont {Chan}}, \bibinfo {author}
  {\bibfnamefont {B.}~\bibnamefont {Cousins}}, \bibinfo {author} {\bibfnamefont
  {J.~D.~E.}\ \bibnamefont {Creighton}}, \bibinfo {author} {\bibfnamefont
  {B.}~\bibnamefont {Ewing}}, \bibinfo {author} {\bibfnamefont
  {H.}~\bibnamefont {Fong}}, \bibinfo {author} {\bibfnamefont {P.}~\bibnamefont
  {Godwin}}, \bibinfo {author} {\bibfnamefont {C.}~\bibnamefont {Hanna}},
  \bibinfo {author} {\bibfnamefont {S.}~\bibnamefont {Hooper}}, \bibinfo
  {author} {\bibfnamefont {R.}~\bibnamefont {Huxford}}, \bibinfo {author}
  {\bibfnamefont {R.}~\bibnamefont {Magee}}, \bibinfo {author} {\bibfnamefont
  {D.}~\bibnamefont {Meacher}}, \bibinfo {author} {\bibfnamefont
  {C.}~\bibnamefont {Messick}}, \bibinfo {author} {\bibfnamefont
  {S.}~\bibnamefont {Morisaki}}, \bibinfo {author} {\bibfnamefont
  {D.}~\bibnamefont {Mukherjee}}, \bibinfo {author} {\bibfnamefont
  {H.}~\bibnamefont {Ohta}}, \bibinfo {author} {\bibfnamefont {A.}~\bibnamefont
  {Pace}}, \bibinfo {author} {\bibfnamefont {S.}~\bibnamefont {Privitera}},
  \bibinfo {author} {\bibfnamefont {I.}~\bibnamefont {de~Ruiter}}, \bibinfo
  {author} {\bibfnamefont {S.}~\bibnamefont {Sachdev}}, \bibinfo {author}
  {\bibfnamefont {L.}~\bibnamefont {Singer}}, \bibinfo {author} {\bibfnamefont
  {D.}~\bibnamefont {Singh}}, \bibinfo {author} {\bibfnamefont
  {R.}~\bibnamefont {Tapia}}, \bibinfo {author} {\bibfnamefont
  {L.}~\bibnamefont {Tsukada}}, \bibinfo {author} {\bibfnamefont
  {D.}~\bibnamefont {Tsuna}}, \bibinfo {author} {\bibfnamefont
  {T.}~\bibnamefont {Tsutsui}}, \bibinfo {author} {\bibfnamefont
  {K.}~\bibnamefont {Ueno}}, \bibinfo {author} {\bibfnamefont {A.}~\bibnamefont
  {Viets}}, \bibinfo {author} {\bibfnamefont {L.}~\bibnamefont {Wade}},\ and\
  \bibinfo {author} {\bibfnamefont {M.}~\bibnamefont {Wade}},\ }\href@noop {}
  {\bibinfo {title} {Gstlal: A software framework for gravitational wave
  discovery}} (\bibinfo {year} {2020}),\ \Eprint
  {https://arxiv.org/abs/2010.05082} {arXiv:2010.05082 [astro-ph.IM]}
  \BibitemShut {NoStop}%
\bibitem [{\citenamefont {Nitz}\ \emph {et~al.}(2021)\citenamefont {Nitz},
  \citenamefont {Capano}, \citenamefont {Kumar}, \citenamefont {Wang},
  \citenamefont {Kastha}, \citenamefont {Schäfer}, \citenamefont {Dhurkunde},\
  and\ \citenamefont {Cabero}}]{Nitz_2021}%
  \BibitemOpen
  \bibfield  {author} {\bibinfo {author} {\bibfnamefont {A.~H.}\ \bibnamefont
  {Nitz}}, \bibinfo {author} {\bibfnamefont {C.~D.}\ \bibnamefont {Capano}},
  \bibinfo {author} {\bibfnamefont {S.}~\bibnamefont {Kumar}}, \bibinfo
  {author} {\bibfnamefont {Y.-F.}\ \bibnamefont {Wang}}, \bibinfo {author}
  {\bibfnamefont {S.}~\bibnamefont {Kastha}}, \bibinfo {author} {\bibfnamefont
  {M.}~\bibnamefont {Schäfer}}, \bibinfo {author} {\bibfnamefont
  {R.}~\bibnamefont {Dhurkunde}},\ and\ \bibinfo {author} {\bibfnamefont
  {M.}~\bibnamefont {Cabero}},\ }\href
  {https://doi.org/10.3847/1538-4357/ac1c03} {\bibfield  {journal} {\bibinfo
  {journal} {The Astrophysical Journal}\ }\textbf {\bibinfo {volume} {922}},\
  \bibinfo {pages} {76} (\bibinfo {year} {2021})}\BibitemShut {NoStop}%
\bibitem [{\citenamefont {Aasi}\ \emph
  {et~al.}(2015{\natexlab{a}})\citenamefont {Aasi}, \citenamefont {Abbott},
  \citenamefont {Abbott}, \citenamefont {Abbott}, \citenamefont {Abernathy},
  \citenamefont {Ackley}, \citenamefont {Adams}, \citenamefont {Adams},
  \citenamefont {Addesso}, \citenamefont {Adhikari} \emph
  {et~al.}}]{aligo2015}%
  \BibitemOpen
  \bibfield  {author} {\bibinfo {author} {\bibfnamefont {J.}~\bibnamefont
  {Aasi}}, \bibinfo {author} {\bibfnamefont {B.~P.}\ \bibnamefont {Abbott}},
  \bibinfo {author} {\bibfnamefont {R.}~\bibnamefont {Abbott}}, \bibinfo
  {author} {\bibfnamefont {T.}~\bibnamefont {Abbott}}, \bibinfo {author}
  {\bibfnamefont {M.~R.}\ \bibnamefont {Abernathy}}, \bibinfo {author}
  {\bibfnamefont {K.}~\bibnamefont {Ackley}}, \bibinfo {author} {\bibfnamefont
  {C.}~\bibnamefont {Adams}}, \bibinfo {author} {\bibfnamefont
  {T.}~\bibnamefont {Adams}}, \bibinfo {author} {\bibfnamefont
  {P.}~\bibnamefont {Addesso}}, \bibinfo {author} {\bibfnamefont {R.~X.}\
  \bibnamefont {Adhikari}}, \emph {et~al.},\ }\href
  {https://doi.org/10.1088/0264-9381/32/7/074001} {\bibfield  {journal}
  {\bibinfo  {journal} {Classical and Quantum Gravity}\ }\textbf {\bibinfo
  {volume} {32}},\ \bibinfo {pages} {074001} (\bibinfo {year}
  {2015}{\natexlab{a}})}\BibitemShut {NoStop}%
\bibitem [{\citenamefont {Acernese}\ \emph {et~al.}(2014)\citenamefont
  {Acernese}, \citenamefont {Agathos}, \citenamefont {Agatsuma}, \citenamefont
  {Aisa}, \citenamefont {Allemandou}, \citenamefont {Allocca}, \citenamefont
  {Amarni}, \citenamefont {Astone}, \citenamefont {Balestri}, \citenamefont
  {Ballardin}, \citenamefont {Barone}, \citenamefont {Baronick} \emph
  {et~al.}}]{Acernese_2014}%
  \BibitemOpen
  \bibfield  {author} {\bibinfo {author} {\bibfnamefont {F.}~\bibnamefont
  {Acernese}}, \bibinfo {author} {\bibfnamefont {M.}~\bibnamefont {Agathos}},
  \bibinfo {author} {\bibfnamefont {K.}~\bibnamefont {Agatsuma}}, \bibinfo
  {author} {\bibfnamefont {D.}~\bibnamefont {Aisa}}, \bibinfo {author}
  {\bibfnamefont {N.}~\bibnamefont {Allemandou}}, \bibinfo {author}
  {\bibfnamefont {A.}~\bibnamefont {Allocca}}, \bibinfo {author} {\bibfnamefont
  {J.}~\bibnamefont {Amarni}}, \bibinfo {author} {\bibfnamefont
  {P.}~\bibnamefont {Astone}}, \bibinfo {author} {\bibfnamefont
  {G.}~\bibnamefont {Balestri}}, \bibinfo {author} {\bibfnamefont
  {G.}~\bibnamefont {Ballardin}}, \bibinfo {author} {\bibfnamefont
  {F.}~\bibnamefont {Barone}}, \bibinfo {author} {\bibfnamefont {J.-P.}\
  \bibnamefont {Baronick}}, \emph {et~al.},\ }\href
  {https://doi.org/10.1088/0264-9381/32/2/024001} {\bibfield  {journal}
  {\bibinfo  {journal} {Classical and Quantum Gravity}\ }\textbf {\bibinfo
  {volume} {32}},\ \bibinfo {pages} {024001} (\bibinfo {year}
  {2014})}\BibitemShut {NoStop}%
\bibitem [{\citenamefont {Abbott}\ \emph {et~al.}(2019)\citenamefont {Abbott},
  \citenamefont {Abbott}, \citenamefont {Abbott}, \citenamefont {Abraham},
  \citenamefont {Acernese}, \citenamefont {Ackley}, \citenamefont {Adams},
  \citenamefont {Adhikari}, \citenamefont {Adya}, \citenamefont {Affeldt} \emph
  {et~al.}}]{gwtc1}%
  \BibitemOpen
  \bibfield  {author} {\bibinfo {author} {\bibfnamefont {B.}~\bibnamefont
  {Abbott}}, \bibinfo {author} {\bibfnamefont {R.}~\bibnamefont {Abbott}},
  \bibinfo {author} {\bibfnamefont {T.}~\bibnamefont {Abbott}}, \bibinfo
  {author} {\bibfnamefont {S.}~\bibnamefont {Abraham}}, \bibinfo {author}
  {\bibfnamefont {F.}~\bibnamefont {Acernese}}, \bibinfo {author}
  {\bibfnamefont {K.}~\bibnamefont {Ackley}}, \bibinfo {author} {\bibfnamefont
  {C.}~\bibnamefont {Adams}}, \bibinfo {author} {\bibfnamefont
  {R.}~\bibnamefont {Adhikari}}, \bibinfo {author} {\bibfnamefont
  {V.}~\bibnamefont {Adya}}, \bibinfo {author} {\bibfnamefont {C.}~\bibnamefont
  {Affeldt}}, \emph {et~al.},\ }\bibfield  {journal} {\bibinfo  {journal}
  {Physical Review X}\ }\textbf {\bibinfo {volume} {9}},\ \href
  {https://doi.org/10.1103/physrevx.9.031040} {10.1103/physrevx.9.031040}
  (\bibinfo {year} {2019})\BibitemShut {NoStop}%
\bibitem [{\citenamefont {Abbott}\ \emph
  {et~al.}(2021{\natexlab{a}})\citenamefont {Abbott}, \citenamefont {Abbott},
  \citenamefont {Abraham}, \citenamefont {Acernese}, \citenamefont {Ackley},
  \citenamefont {Adams}, \citenamefont {Adams}, \citenamefont {Adhikari},
  \citenamefont {Adya}, \citenamefont {Affeldt}, \citenamefont {Agathos},
  \citenamefont {Agatsuma}, \citenamefont {Aggarwal}, \citenamefont {Aguiar},
  \citenamefont {Aiello}, \citenamefont {Ain}, \citenamefont {Ajith},
  \citenamefont {Akcay}, \citenamefont {Allen}, \citenamefont {Allocca},
  \citenamefont {Altin}, \citenamefont {Amato}, \citenamefont {Anand} \emph
  {et~al.}}]{gwtc2}%
  \BibitemOpen
  \bibfield  {author} {\bibinfo {author} {\bibfnamefont {R.}~\bibnamefont
  {Abbott}}, \bibinfo {author} {\bibfnamefont {T.}~\bibnamefont {Abbott}},
  \bibinfo {author} {\bibfnamefont {S.}~\bibnamefont {Abraham}}, \bibinfo
  {author} {\bibfnamefont {F.}~\bibnamefont {Acernese}}, \bibinfo {author}
  {\bibfnamefont {K.}~\bibnamefont {Ackley}}, \bibinfo {author} {\bibfnamefont
  {A.}~\bibnamefont {Adams}}, \bibinfo {author} {\bibfnamefont
  {C.}~\bibnamefont {Adams}}, \bibinfo {author} {\bibfnamefont
  {R.}~\bibnamefont {Adhikari}}, \bibinfo {author} {\bibfnamefont
  {V.}~\bibnamefont {Adya}}, \bibinfo {author} {\bibfnamefont {C.}~\bibnamefont
  {Affeldt}}, \bibinfo {author} {\bibfnamefont {M.}~\bibnamefont {Agathos}},
  \bibinfo {author} {\bibfnamefont {K.}~\bibnamefont {Agatsuma}}, \bibinfo
  {author} {\bibfnamefont {N.}~\bibnamefont {Aggarwal}}, \bibinfo {author}
  {\bibfnamefont {O.}~\bibnamefont {Aguiar}}, \bibinfo {author} {\bibfnamefont
  {L.}~\bibnamefont {Aiello}}, \bibinfo {author} {\bibfnamefont
  {A.}~\bibnamefont {Ain}}, \bibinfo {author} {\bibfnamefont {P.}~\bibnamefont
  {Ajith}}, \bibinfo {author} {\bibfnamefont {S.}~\bibnamefont {Akcay}},
  \bibinfo {author} {\bibfnamefont {G.}~\bibnamefont {Allen}}, \bibinfo
  {author} {\bibfnamefont {A.}~\bibnamefont {Allocca}}, \bibinfo {author}
  {\bibfnamefont {P.}~\bibnamefont {Altin}}, \bibinfo {author} {\bibfnamefont
  {A.}~\bibnamefont {Amato}}, \bibinfo {author} {\bibfnamefont
  {S.}~\bibnamefont {Anand}}, \emph {et~al.},\ }\bibfield  {journal} {\bibinfo
  {journal} {Physical Review X}\ }\textbf {\bibinfo {volume} {11}},\ \href
  {https://doi.org/10.1103/physrevx.11.021053} {10.1103/physrevx.11.021053}
  (\bibinfo {year} {2021}{\natexlab{a}})\BibitemShut {NoStop}%
\bibitem [{\citenamefont {{The LIGO Scientific Collaboration}}\ \emph
  {et~al.}(2021{\natexlab{a}})\citenamefont {{The LIGO Scientific
  Collaboration}}, \citenamefont {{The Virgo Collaboration}}, \citenamefont
  {Abbott}, \citenamefont {Abbott}, \citenamefont {Acernese}, \citenamefont
  {Ackley}, \citenamefont {Adams}, \citenamefont {Adhikari}, \citenamefont
  {Adhikari}, \citenamefont {Adya}, \citenamefont {Affeldt}, \citenamefont
  {Agarwal}, \citenamefont {Agathos}, \citenamefont {Agatsuma} \emph
  {et~al.}}]{gwtc21}%
  \BibitemOpen
  \bibfield  {author} {\bibinfo {author} {\bibnamefont {{The LIGO Scientific
  Collaboration}}}, \bibinfo {author} {\bibnamefont {{The Virgo
  Collaboration}}}, \bibinfo {author} {\bibfnamefont {R.}~\bibnamefont
  {Abbott}}, \bibinfo {author} {\bibfnamefont {T.~D.}\ \bibnamefont {Abbott}},
  \bibinfo {author} {\bibfnamefont {F.}~\bibnamefont {Acernese}}, \bibinfo
  {author} {\bibfnamefont {K.}~\bibnamefont {Ackley}}, \bibinfo {author}
  {\bibfnamefont {C.}~\bibnamefont {Adams}}, \bibinfo {author} {\bibfnamefont
  {N.}~\bibnamefont {Adhikari}}, \bibinfo {author} {\bibfnamefont {R.~X.}\
  \bibnamefont {Adhikari}}, \bibinfo {author} {\bibfnamefont {V.~B.}\
  \bibnamefont {Adya}}, \bibinfo {author} {\bibfnamefont {C.}~\bibnamefont
  {Affeldt}}, \bibinfo {author} {\bibfnamefont {D.}~\bibnamefont {Agarwal}},
  \bibinfo {author} {\bibfnamefont {M.}~\bibnamefont {Agathos}}, \bibinfo
  {author} {\bibfnamefont {K.}~\bibnamefont {Agatsuma}}, \emph {et~al.},\
  }\href {https://doi.org/10.48550/ARXIV.2108.01045} {\bibinfo {title}
  {Gwtc-2.1: Deep extended catalog of compact binary coalescences observed by
  ligo and virgo during the first half of the third observing run}} (\bibinfo
  {year} {2021}{\natexlab{a}})\BibitemShut {NoStop}%
\bibitem [{\citenamefont {{The LIGO Scientific Collaboration}}\ \emph
  {et~al.}(2021{\natexlab{b}})\citenamefont {{The LIGO Scientific
  Collaboration}}, \citenamefont {{The Virgo Collaboration}}, \citenamefont
  {{The KAGRA Collaboration}}, \citenamefont {Abbott}, \citenamefont {Abbott},
  \citenamefont {Acernese}, \citenamefont {Ackley}, \citenamefont {Adams},
  \citenamefont {Adhikari}, \citenamefont {Adhikari}, \citenamefont {Adya},
  \citenamefont {Affeldt}, \citenamefont {Agarwal}, \citenamefont {Agathos},
  \citenamefont {Agatsuma} \emph {et~al.}}]{gwtc3}%
  \BibitemOpen
  \bibfield  {author} {\bibinfo {author} {\bibnamefont {{The LIGO Scientific
  Collaboration}}}, \bibinfo {author} {\bibnamefont {{The Virgo
  Collaboration}}}, \bibinfo {author} {\bibnamefont {{The KAGRA
  Collaboration}}}, \bibinfo {author} {\bibfnamefont {R.}~\bibnamefont
  {Abbott}}, \bibinfo {author} {\bibfnamefont {T.~D.}\ \bibnamefont {Abbott}},
  \bibinfo {author} {\bibfnamefont {F.}~\bibnamefont {Acernese}}, \bibinfo
  {author} {\bibfnamefont {K.}~\bibnamefont {Ackley}}, \bibinfo {author}
  {\bibfnamefont {C.}~\bibnamefont {Adams}}, \bibinfo {author} {\bibfnamefont
  {N.}~\bibnamefont {Adhikari}}, \bibinfo {author} {\bibfnamefont {R.~X.}\
  \bibnamefont {Adhikari}}, \bibinfo {author} {\bibfnamefont {V.~B.}\
  \bibnamefont {Adya}}, \bibinfo {author} {\bibfnamefont {C.}~\bibnamefont
  {Affeldt}}, \bibinfo {author} {\bibfnamefont {D.}~\bibnamefont {Agarwal}},
  \bibinfo {author} {\bibfnamefont {M.}~\bibnamefont {Agathos}}, \bibinfo
  {author} {\bibfnamefont {K.}~\bibnamefont {Agatsuma}}, \emph {et~al.},\
  }\href {https://doi.org/10.48550/ARXIV.2111.03606} {\bibinfo {title} {Gwtc-3:
  Compact binary coalescences observed by ligo and virgo during the second part
  of the third observing run}} (\bibinfo {year}
  {2021}{\natexlab{b}})\BibitemShut {NoStop}%
\bibitem [{\citenamefont {Miyamoto}\ \emph {et~al.}(2022)\citenamefont
  {Miyamoto}, \citenamefont {Morr\'as}, \citenamefont {Yamamoto}, \citenamefont
  {Kuroyanagi},\ and\ \citenamefont {Nesseris}}]{miyamoto2022gravitational}%
  \BibitemOpen
  \bibfield  {author} {\bibinfo {author} {\bibfnamefont {K.}~\bibnamefont
  {Miyamoto}}, \bibinfo {author} {\bibfnamefont {G.}~\bibnamefont {Morr\'as}},
  \bibinfo {author} {\bibfnamefont {T.~S.}\ \bibnamefont {Yamamoto}}, \bibinfo
  {author} {\bibfnamefont {S.}~\bibnamefont {Kuroyanagi}},\ and\ \bibinfo
  {author} {\bibfnamefont {S.}~\bibnamefont {Nesseris}},\ }\href
  {https://doi.org/10.1103/PhysRevResearch.4.033150} {\bibfield  {journal}
  {\bibinfo  {journal} {Phys. Rev. Res.}\ }\textbf {\bibinfo {volume} {4}},\
  \bibinfo {pages} {033150} (\bibinfo {year} {2022})}\BibitemShut {NoStop}%
\bibitem [{\citenamefont {Gao}\ \emph {et~al.}(2022)\citenamefont {Gao},
  \citenamefont {Hayes}, \citenamefont {Croke}, \citenamefont {Messenger},\
  and\ \citenamefont {Veitch}}]{PhysRevResearch.4.023006}%
  \BibitemOpen
  \bibfield  {author} {\bibinfo {author} {\bibfnamefont {S.}~\bibnamefont
  {Gao}}, \bibinfo {author} {\bibfnamefont {F.}~\bibnamefont {Hayes}}, \bibinfo
  {author} {\bibfnamefont {S.}~\bibnamefont {Croke}}, \bibinfo {author}
  {\bibfnamefont {C.}~\bibnamefont {Messenger}},\ and\ \bibinfo {author}
  {\bibfnamefont {J.}~\bibnamefont {Veitch}},\ }\href
  {https://doi.org/10.1103/PhysRevResearch.4.023006} {\bibfield  {journal}
  {\bibinfo  {journal} {Phys. Rev. Research}\ }\textbf {\bibinfo {volume}
  {4}},\ \bibinfo {pages} {023006} (\bibinfo {year} {2022})}\BibitemShut
  {NoStop}%
\bibitem [{\citenamefont {Escrig}\ \emph {et~al.}(2023)\citenamefont {Escrig},
  \citenamefont {Campos}, \citenamefont {Casares},\ and\ \citenamefont
  {Martin-Delgado}}]{Escrig_2023}%
  \BibitemOpen
  \bibfield  {author} {\bibinfo {author} {\bibfnamefont {G.}~\bibnamefont
  {Escrig}}, \bibinfo {author} {\bibfnamefont {R.}~\bibnamefont {Campos}},
  \bibinfo {author} {\bibfnamefont {P.~A.~M.}\ \bibnamefont {Casares}},\ and\
  \bibinfo {author} {\bibfnamefont {M.~A.}\ \bibnamefont {Martin-Delgado}},\
  }\href {https://doi.org/10.1088/1361-6382/acafcf} {\bibfield  {journal}
  {\bibinfo  {journal} {Classical and Quantum Gravity}\ }\textbf {\bibinfo
  {volume} {40}},\ \bibinfo {pages} {045001} (\bibinfo {year}
  {2023})}\BibitemShut {NoStop}%
\bibitem [{\citenamefont {Zhang}\ \emph {et~al.}(2022)\citenamefont {Zhang},
  \citenamefont {Li},\ and\ \citenamefont
  {Yuan}}]{https://doi.org/10.48550/arxiv.2201.11495}%
  \BibitemOpen
  \bibfield  {author} {\bibinfo {author} {\bibfnamefont {X.-M.}\ \bibnamefont
  {Zhang}}, \bibinfo {author} {\bibfnamefont {T.}~\bibnamefont {Li}},\ and\
  \bibinfo {author} {\bibfnamefont {X.}~\bibnamefont {Yuan}},\ }\href
  {https://doi.org/10.48550/ARXIV.2201.11495} {\bibinfo {title} {Quantum state
  preparation with optimal circuit depth: Implementations and applications}}
  (\bibinfo {year} {2022})\BibitemShut {NoStop}%
\bibitem [{\citenamefont {Buhrman}\ \emph {et~al.}(2001)\citenamefont
  {Buhrman}, \citenamefont {Cleve}, \citenamefont {Watrous},\ and\
  \citenamefont {de~Wolf}}]{PhysRevLett.87.167902}%
  \BibitemOpen
  \bibfield  {author} {\bibinfo {author} {\bibfnamefont {H.}~\bibnamefont
  {Buhrman}}, \bibinfo {author} {\bibfnamefont {R.}~\bibnamefont {Cleve}},
  \bibinfo {author} {\bibfnamefont {J.}~\bibnamefont {Watrous}},\ and\ \bibinfo
  {author} {\bibfnamefont {R.}~\bibnamefont {de~Wolf}},\ }\href
  {https://doi.org/10.1103/PhysRevLett.87.167902} {\bibfield  {journal}
  {\bibinfo  {journal} {Phys. Rev. Lett.}\ }\textbf {\bibinfo {volume} {87}},\
  \bibinfo {pages} {167902} (\bibinfo {year} {2001})}\BibitemShut {NoStop}%
\bibitem [{\citenamefont {Kang}\ \emph {et~al.}(2019)\citenamefont {Kang},
  \citenamefont {Heo}, \citenamefont {Choi}, \citenamefont {Moon},\ and\
  \citenamefont {Han}}]{kang2019implementation}%
  \BibitemOpen
  \bibfield  {author} {\bibinfo {author} {\bibfnamefont {M.-S.}\ \bibnamefont
  {Kang}}, \bibinfo {author} {\bibfnamefont {J.}~\bibnamefont {Heo}}, \bibinfo
  {author} {\bibfnamefont {S.-G.}\ \bibnamefont {Choi}}, \bibinfo {author}
  {\bibfnamefont {S.}~\bibnamefont {Moon}},\ and\ \bibinfo {author}
  {\bibfnamefont {S.-W.}\ \bibnamefont {Han}},\ }\href@noop {} {\bibfield
  {journal} {\bibinfo  {journal} {Scientific reports}\ }\textbf {\bibinfo
  {volume} {9}},\ \bibinfo {pages} {1} (\bibinfo {year} {2019})}\BibitemShut
  {NoStop}%
\bibitem [{\citenamefont {Cincio}\ \emph {et~al.}(2018)\citenamefont {Cincio},
  \citenamefont {Suba{\c{s}}{\i}}, \citenamefont {Sornborger},\ and\
  \citenamefont {Coles}}]{Cincio_2018}%
  \BibitemOpen
  \bibfield  {author} {\bibinfo {author} {\bibfnamefont {L.}~\bibnamefont
  {Cincio}}, \bibinfo {author} {\bibfnamefont {Y.}~\bibnamefont
  {Suba{\c{s}}{\i}}}, \bibinfo {author} {\bibfnamefont {A.~T.}\ \bibnamefont
  {Sornborger}},\ and\ \bibinfo {author} {\bibfnamefont {P.~J.}\ \bibnamefont
  {Coles}},\ }\href {https://doi.org/10.1088/1367-2630/aae94a} {\bibfield
  {journal} {\bibinfo  {journal} {New Journal of Physics}\ }\textbf {\bibinfo
  {volume} {20}},\ \bibinfo {pages} {113022} (\bibinfo {year}
  {2018})}\BibitemShut {NoStop}%
\bibitem [{\citenamefont {Treinish}\ \emph {et~al.}(2022)\citenamefont
  {Treinish}, \citenamefont {Gambetta}, \citenamefont {Nation}, \citenamefont
  {Kassebaum}, \citenamefont {qiskit bot}, \citenamefont {Rodríguez},
  \citenamefont {González}, \citenamefont {Hu}, \citenamefont {Krsulich},
  \citenamefont {Zdanski}, \citenamefont {Garrison}, \citenamefont {Yu},
  \citenamefont {Gacon}, \citenamefont {McKay}, \citenamefont {Gomez},
  \citenamefont {Capelluto}, \citenamefont {Travis-S-IBM}, \citenamefont
  {Marques}, \citenamefont {Panigrahi}, \citenamefont {Lishman}, \citenamefont
  {lerongil}, \citenamefont {Rahman}, \citenamefont {Wood}, \citenamefont
  {Bello}, \citenamefont {Itoko}, \citenamefont {Singh}, \citenamefont {Drew},
  \citenamefont {Arbel}, \citenamefont {Schwarm},\ and\ \citenamefont
  {Daniel}}]{treinish_qiskit_2022}%
  \BibitemOpen
  \bibfield  {author} {\bibinfo {author} {\bibfnamefont {M.}~\bibnamefont
  {Treinish}}, \bibinfo {author} {\bibfnamefont {J.}~\bibnamefont {Gambetta}},
  \bibinfo {author} {\bibfnamefont {P.}~\bibnamefont {Nation}}, \bibinfo
  {author} {\bibfnamefont {P.}~\bibnamefont {Kassebaum}}, \bibinfo {author}
  {\bibnamefont {qiskit bot}}, \bibinfo {author} {\bibfnamefont {D.~M.}\
  \bibnamefont {Rodríguez}}, \bibinfo {author} {\bibfnamefont {S.~d. l.~P.}\
  \bibnamefont {González}}, \bibinfo {author} {\bibfnamefont {S.}~\bibnamefont
  {Hu}}, \bibinfo {author} {\bibfnamefont {K.}~\bibnamefont {Krsulich}},
  \bibinfo {author} {\bibfnamefont {L.}~\bibnamefont {Zdanski}}, \bibinfo
  {author} {\bibfnamefont {J.}~\bibnamefont {Garrison}}, \bibinfo {author}
  {\bibfnamefont {J.}~\bibnamefont {Yu}}, \bibinfo {author} {\bibfnamefont
  {J.}~\bibnamefont {Gacon}}, \bibinfo {author} {\bibfnamefont
  {D.}~\bibnamefont {McKay}}, \bibinfo {author} {\bibfnamefont
  {J.}~\bibnamefont {Gomez}}, \bibinfo {author} {\bibfnamefont
  {L.}~\bibnamefont {Capelluto}}, \bibinfo {author} {\bibnamefont
  {Travis-S-IBM}}, \bibinfo {author} {\bibfnamefont {M.}~\bibnamefont
  {Marques}}, \bibinfo {author} {\bibfnamefont {A.}~\bibnamefont {Panigrahi}},
  \bibinfo {author} {\bibfnamefont {J.}~\bibnamefont {Lishman}}, \bibinfo
  {author} {\bibnamefont {lerongil}}, \bibinfo {author} {\bibfnamefont {R.~I.}\
  \bibnamefont {Rahman}}, \bibinfo {author} {\bibfnamefont {S.}~\bibnamefont
  {Wood}}, \bibinfo {author} {\bibfnamefont {L.}~\bibnamefont {Bello}},
  \bibinfo {author} {\bibfnamefont {T.}~\bibnamefont {Itoko}}, \bibinfo
  {author} {\bibfnamefont {D.}~\bibnamefont {Singh}}, \bibinfo {author}
  {\bibnamefont {Drew}}, \bibinfo {author} {\bibfnamefont {E.}~\bibnamefont
  {Arbel}}, \bibinfo {author} {\bibfnamefont {J.}~\bibnamefont {Schwarm}},\
  and\ \bibinfo {author} {\bibfnamefont {J.}~\bibnamefont {Daniel}},\ }\href
  {https://doi.org/10.5281/zenodo.6403335} {\bibinfo {title} {Qiskit: {An}
  {Open}-source {Framework} for {Quantum} {Computing}}} (\bibinfo {year}
  {2022})\BibitemShut {NoStop}%
\bibitem [{\citenamefont {Abbott}\ \emph
  {et~al.}(2020{\natexlab{a}})\citenamefont {Abbott}, \citenamefont {Abbott},
  \citenamefont {Abraham}, \citenamefont {Acernese}, \citenamefont {Ackley},
  \citenamefont {Adams}, \citenamefont {Adhikari} \emph
  {et~al.}}]{PhysRevLett.125.101102}%
  \BibitemOpen
  \bibfield  {author} {\bibinfo {author} {\bibfnamefont {R.}~\bibnamefont
  {Abbott}}, \bibinfo {author} {\bibfnamefont {T.~D.}\ \bibnamefont {Abbott}},
  \bibinfo {author} {\bibfnamefont {S.}~\bibnamefont {Abraham}}, \bibinfo
  {author} {\bibfnamefont {F.}~\bibnamefont {Acernese}}, \bibinfo {author}
  {\bibfnamefont {K.}~\bibnamefont {Ackley}}, \bibinfo {author} {\bibfnamefont
  {C.}~\bibnamefont {Adams}}, \bibinfo {author} {\bibfnamefont {R.~X.}\
  \bibnamefont {Adhikari}}, \emph {et~al.} (\bibinfo {collaboration} {LIGO
  Scientific Collaboration and Virgo Collaboration}),\ }\href
  {https://doi.org/10.1103/PhysRevLett.125.101102} {\bibfield  {journal}
  {\bibinfo  {journal} {Phys. Rev. Lett.}\ }\textbf {\bibinfo {volume} {125}},\
  \bibinfo {pages} {101102} (\bibinfo {year} {2020}{\natexlab{a}})}\BibitemShut
  {NoStop}%
\bibitem [{\citenamefont {Koch}\ \emph {et~al.}(2007)\citenamefont {Koch},
  \citenamefont {Yu}, \citenamefont {Gambetta}, \citenamefont {Houck},
  \citenamefont {Schuster}, \citenamefont {Majer}, \citenamefont {Blais},
  \citenamefont {Devoret}, \citenamefont {Girvin},\ and\ \citenamefont
  {Schoelkopf}}]{Koch2007}%
  \BibitemOpen
  \bibfield  {author} {\bibinfo {author} {\bibfnamefont {J.}~\bibnamefont
  {Koch}}, \bibinfo {author} {\bibfnamefont {T.~M.}\ \bibnamefont {Yu}},
  \bibinfo {author} {\bibfnamefont {J.~M.}\ \bibnamefont {Gambetta}}, \bibinfo
  {author} {\bibfnamefont {A.~A.}\ \bibnamefont {Houck}}, \bibinfo {author}
  {\bibfnamefont {D.~I.}\ \bibnamefont {Schuster}}, \bibinfo {author}
  {\bibfnamefont {J.}~\bibnamefont {Majer}}, \bibinfo {author} {\bibfnamefont
  {A.}~\bibnamefont {Blais}}, \bibinfo {author} {\bibfnamefont {M.~H.}\
  \bibnamefont {Devoret}}, \bibinfo {author} {\bibfnamefont {S.~M.}\
  \bibnamefont {Girvin}},\ and\ \bibinfo {author} {\bibfnamefont {R.~J.}\
  \bibnamefont {Schoelkopf}},\ }\href
  {https://doi.org/10.1103/PhysRevA.76.042319} {\bibfield  {journal} {\bibinfo
  {journal} {Physical Review A}\ }\textbf {\bibinfo {volume} {76}},\ \bibinfo
  {pages} {042319} (\bibinfo {year} {2007})}\BibitemShut {NoStop}%
\bibitem [{\citenamefont {Chamberland}\ \emph {et~al.}(2019)\citenamefont
  {Chamberland}, \citenamefont {Zhu}, \citenamefont {Yoder}, \citenamefont
  {Hertzberg},\ and\ \citenamefont {Cross}}]{Chamberland2019}%
  \BibitemOpen
  \bibfield  {author} {\bibinfo {author} {\bibfnamefont {C.}~\bibnamefont
  {Chamberland}}, \bibinfo {author} {\bibfnamefont {G.}~\bibnamefont {Zhu}},
  \bibinfo {author} {\bibfnamefont {T.~J.}\ \bibnamefont {Yoder}}, \bibinfo
  {author} {\bibfnamefont {J.~B.}\ \bibnamefont {Hertzberg}},\ and\ \bibinfo
  {author} {\bibfnamefont {A.~W.}\ \bibnamefont {Cross}},\ }\href
  {https://doi.org/10.1103/PhysRevX.10.011022} {\bibfield  {journal} {\bibinfo
  {journal} {Physical Review X}\ }\textbf {\bibinfo {volume} {10}},\ \bibinfo
  {pages} {1} (\bibinfo {year} {2019})},\ \Eprint
  {https://arxiv.org/abs/1907.09528} {1907.09528} \BibitemShut {NoStop}%
\bibitem [{\citenamefont {Chow}\ \emph {et~al.}(2011)\citenamefont {Chow},
  \citenamefont {C{\'{o}}rcoles}, \citenamefont {Gambetta}, \citenamefont
  {Rigetti}, \citenamefont {Johnson}, \citenamefont {Smolin}, \citenamefont
  {Rozen}, \citenamefont {Keefe}, \citenamefont {Rothwell}, \citenamefont
  {Ketchen},\ and\ \citenamefont {Steffen}}]{Chow2011}%
  \BibitemOpen
  \bibfield  {author} {\bibinfo {author} {\bibfnamefont {J.~M.}\ \bibnamefont
  {Chow}}, \bibinfo {author} {\bibfnamefont {A.~D.}\ \bibnamefont
  {C{\'{o}}rcoles}}, \bibinfo {author} {\bibfnamefont {J.~M.}\ \bibnamefont
  {Gambetta}}, \bibinfo {author} {\bibfnamefont {C.}~\bibnamefont {Rigetti}},
  \bibinfo {author} {\bibfnamefont {B.~R.}\ \bibnamefont {Johnson}}, \bibinfo
  {author} {\bibfnamefont {J.~A.}\ \bibnamefont {Smolin}}, \bibinfo {author}
  {\bibfnamefont {J.~R.}\ \bibnamefont {Rozen}}, \bibinfo {author}
  {\bibfnamefont {G.~A.}\ \bibnamefont {Keefe}}, \bibinfo {author}
  {\bibfnamefont {M.~B.}\ \bibnamefont {Rothwell}}, \bibinfo {author}
  {\bibfnamefont {M.~B.}\ \bibnamefont {Ketchen}},\ and\ \bibinfo {author}
  {\bibfnamefont {M.}~\bibnamefont {Steffen}},\ }\href
  {https://doi.org/10.1103/PhysRevLett.107.080502} {\bibfield  {journal}
  {\bibinfo  {journal} {Physical Review Letters}\ }\textbf {\bibinfo {volume}
  {107}},\ \bibinfo {pages} {080502} (\bibinfo {year} {2011})}\BibitemShut
  {NoStop}%
\bibitem [{\citenamefont {Sheldon}\ \emph {et~al.}(2016)\citenamefont
  {Sheldon}, \citenamefont {Magesan}, \citenamefont {Chow},\ and\ \citenamefont
  {Gambetta}}]{Sheldon2016}%
  \BibitemOpen
  \bibfield  {author} {\bibinfo {author} {\bibfnamefont {S.}~\bibnamefont
  {Sheldon}}, \bibinfo {author} {\bibfnamefont {E.}~\bibnamefont {Magesan}},
  \bibinfo {author} {\bibfnamefont {J.~M.}\ \bibnamefont {Chow}},\ and\
  \bibinfo {author} {\bibfnamefont {J.~M.}\ \bibnamefont {Gambetta}},\ }\href
  {https://doi.org/10.1103/PhysRevA.93.060302} {\bibfield  {journal} {\bibinfo
  {journal} {Physical Review A}\ }\textbf {\bibinfo {volume} {93}},\ \bibinfo
  {pages} {060302} (\bibinfo {year} {2016})}\BibitemShut {NoStop}%
\bibitem [{\citenamefont {Blais}\ \emph {et~al.}(2004)\citenamefont {Blais},
  \citenamefont {Huang}, \citenamefont {Wallraff}, \citenamefont {Girvin},\
  and\ \citenamefont {Schoelkopf}}]{Blais2004}%
  \BibitemOpen
  \bibfield  {author} {\bibinfo {author} {\bibfnamefont {A.}~\bibnamefont
  {Blais}}, \bibinfo {author} {\bibfnamefont {R.-S.}\ \bibnamefont {Huang}},
  \bibinfo {author} {\bibfnamefont {A.}~\bibnamefont {Wallraff}}, \bibinfo
  {author} {\bibfnamefont {S.~M.}\ \bibnamefont {Girvin}},\ and\ \bibinfo
  {author} {\bibfnamefont {R.~J.}\ \bibnamefont {Schoelkopf}},\ }\bibfield
  {journal} {\bibinfo  {journal} {Physical Review A}\ }\textbf {\bibinfo
  {volume} {69}},\ \href {https://doi.org/10.1103/PhysRevA.69.062320}
  {10.1103/PhysRevA.69.062320} (\bibinfo {year} {2004})\BibitemShut {NoStop}%
\bibitem [{\citenamefont {Wallraff}\ \emph {et~al.}(2004)\citenamefont
  {Wallraff}, \citenamefont {Schuster}, \citenamefont {Blais}, \citenamefont
  {Frunzio}, \citenamefont {Majer}, \citenamefont {Kumar}, \citenamefont
  {Girvin}, \citenamefont {Schoelkopf},\ and\ \citenamefont
  {Huang}}]{Wallraff2004}%
  \BibitemOpen
  \bibfield  {author} {\bibinfo {author} {\bibfnamefont {A.}~\bibnamefont
  {Wallraff}}, \bibinfo {author} {\bibfnamefont {D.~I.}\ \bibnamefont
  {Schuster}}, \bibinfo {author} {\bibfnamefont {A.}~\bibnamefont {Blais}},
  \bibinfo {author} {\bibfnamefont {L.}~\bibnamefont {Frunzio}}, \bibinfo
  {author} {\bibfnamefont {J.}~\bibnamefont {Majer}}, \bibinfo {author}
  {\bibfnamefont {S.}~\bibnamefont {Kumar}}, \bibinfo {author} {\bibfnamefont
  {S.~M.}\ \bibnamefont {Girvin}}, \bibinfo {author} {\bibfnamefont {R.~J.}\
  \bibnamefont {Schoelkopf}},\ and\ \bibinfo {author} {\bibfnamefont {R.-S.
  R.-S.}\ \bibnamefont {Huang}},\ }\href {https://doi.org/10.1038/nature02851}
  {\bibfield  {journal} {\bibinfo  {journal} {Nature}\ }\textbf {\bibinfo
  {volume} {431}},\ \bibinfo {pages} {162} (\bibinfo {year}
  {2004})}\BibitemShut {NoStop}%
\bibitem [{\citenamefont {McKay}\ \emph {et~al.}(2017)\citenamefont {McKay},
  \citenamefont {Wood}, \citenamefont {Sheldon}, \citenamefont {Chow},\ and\
  \citenamefont {Gambetta}}]{McKay2017}%
  \BibitemOpen
  \bibfield  {author} {\bibinfo {author} {\bibfnamefont {D.~C.}\ \bibnamefont
  {McKay}}, \bibinfo {author} {\bibfnamefont {C.~J.}\ \bibnamefont {Wood}},
  \bibinfo {author} {\bibfnamefont {S.}~\bibnamefont {Sheldon}}, \bibinfo
  {author} {\bibfnamefont {J.~M.}\ \bibnamefont {Chow}},\ and\ \bibinfo
  {author} {\bibfnamefont {J.~M.}\ \bibnamefont {Gambetta}},\ }\href
  {https://doi.org/10.1103/PhysRevA.96.022330} {\bibfield  {journal} {\bibinfo
  {journal} {Physical Review A}\ }\textbf {\bibinfo {volume} {96}},\ \bibinfo
  {pages} {1} (\bibinfo {year} {2017})}\BibitemShut {NoStop}%
\bibitem [{\citenamefont {Gambetta}\ \emph {et~al.}(2017)\citenamefont
  {Gambetta}, \citenamefont {Chow},\ and\ \citenamefont
  {Steffen}}]{Gambetta_2017}%
  \BibitemOpen
  \bibfield  {author} {\bibinfo {author} {\bibfnamefont {J.~M.}\ \bibnamefont
  {Gambetta}}, \bibinfo {author} {\bibfnamefont {J.~M.}\ \bibnamefont {Chow}},\
  and\ \bibinfo {author} {\bibfnamefont {M.}~\bibnamefont {Steffen}},\
  }\bibfield  {journal} {\bibinfo  {journal} {npj Quantum Information}\
  }\textbf {\bibinfo {volume} {3}},\ \href
  {https://doi.org/10.1038/s41534-016-0004-0} {10.1038/s41534-016-0004-0}
  (\bibinfo {year} {2017})\BibitemShut {NoStop}%
\bibitem [{\citenamefont {Bravyi}\ \emph {et~al.}(2022)\citenamefont {Bravyi},
  \citenamefont {Dial}, \citenamefont {Gambetta}, \citenamefont {Gil},\ and\
  \citenamefont {Nazario}}]{Bravyi_2022}%
  \BibitemOpen
  \bibfield  {author} {\bibinfo {author} {\bibfnamefont {S.}~\bibnamefont
  {Bravyi}}, \bibinfo {author} {\bibfnamefont {O.}~\bibnamefont {Dial}},
  \bibinfo {author} {\bibfnamefont {J.~M.}\ \bibnamefont {Gambetta}}, \bibinfo
  {author} {\bibfnamefont {D.}~\bibnamefont {Gil}},\ and\ \bibinfo {author}
  {\bibfnamefont {Z.}~\bibnamefont {Nazario}},\ }\href
  {https://doi.org/10.1063/5.0082975} {\bibfield  {journal} {\bibinfo
  {journal} {Journal of Applied Physics}\ }\textbf {\bibinfo {volume} {132}},\
  \bibinfo {pages} {160902} (\bibinfo {year} {2022})}\BibitemShut {NoStop}%
\bibitem [{IQS()}]{IQS}%
  \BibitemOpen
  \href@noop {} {\bibinfo {title} {{IBM Quantum Compute Resources}}},\ \bibinfo
  {howpublished}
  {\url{https://quantum-computing.ibm.com/services/resources?tab=systems}}\BibitemShut
  {NoStop}%
\bibitem [{\citenamefont {Aasi}\ \emph
  {et~al.}(2015{\natexlab{b}})\citenamefont {Aasi}, \citenamefont {Abbott},
  \citenamefont {Abbott}, \citenamefont {Abbott}, \citenamefont {Abernathy},
  \citenamefont {Ackley}, \citenamefont {Adams}, \citenamefont {Adams},
  \citenamefont {Addesso}, \citenamefont {Adhikari} \emph
  {et~al.}}]{TheLIGOScientific:2014jea}%
  \BibitemOpen
  \bibfield  {author} {\bibinfo {author} {\bibfnamefont {J.}~\bibnamefont
  {Aasi}}, \bibinfo {author} {\bibfnamefont {B.~P.}\ \bibnamefont {Abbott}},
  \bibinfo {author} {\bibfnamefont {R.}~\bibnamefont {Abbott}}, \bibinfo
  {author} {\bibfnamefont {T.}~\bibnamefont {Abbott}}, \bibinfo {author}
  {\bibfnamefont {M.~R.}\ \bibnamefont {Abernathy}}, \bibinfo {author}
  {\bibfnamefont {K.}~\bibnamefont {Ackley}}, \bibinfo {author} {\bibfnamefont
  {C.}~\bibnamefont {Adams}}, \bibinfo {author} {\bibfnamefont
  {T.}~\bibnamefont {Adams}}, \bibinfo {author} {\bibfnamefont
  {P.}~\bibnamefont {Addesso}}, \bibinfo {author} {\bibfnamefont {R.~X.}\
  \bibnamefont {Adhikari}}, \emph {et~al.},\ }\href
  {https://doi.org/10.1088/0264-9381/32/7/074001} {\bibfield  {journal}
  {\bibinfo  {journal} {Classical and Quantum Gravity}\ }\textbf {\bibinfo
  {volume} {32}},\ \bibinfo {pages} {074001} (\bibinfo {year}
  {2015}{\natexlab{b}})}\BibitemShut {NoStop}%
\bibitem [{Note1()}]{Note1}%
  \BibitemOpen
  \bibinfo {note} {\protect \url
  {https://www.gw-openscience.org/eventapi/html/GWTC-2/GW190521/}}\BibitemShut
  {NoStop}%
\bibitem [{\citenamefont {Welch}(1967)}]{1161901}%
  \BibitemOpen
  \bibfield  {author} {\bibinfo {author} {\bibfnamefont {P.}~\bibnamefont
  {Welch}},\ }\href {https://doi.org/10.1109/TAU.1967.1161901} {\bibfield
  {journal} {\bibinfo  {journal} {IEEE Transactions on Audio and
  Electroacoustics}\ }\textbf {\bibinfo {volume} {15}},\ \bibinfo {pages} {70}
  (\bibinfo {year} {1967})}\BibitemShut {NoStop}%
\bibitem [{\citenamefont {Hunter}(2007)}]{4160265}%
  \BibitemOpen
  \bibfield  {author} {\bibinfo {author} {\bibfnamefont {J.~D.}\ \bibnamefont
  {Hunter}},\ }\href {https://doi.org/10.1109/MCSE.2007.55} {\bibfield
  {journal} {\bibinfo  {journal} {Computing in Science \& Engineering}\
  }\textbf {\bibinfo {volume} {9}},\ \bibinfo {pages} {90} (\bibinfo {year}
  {2007})}\BibitemShut {NoStop}%
\bibitem [{\citenamefont {Abbott}\ \emph
  {et~al.}(2020{\natexlab{b}})\citenamefont {Abbott}, \citenamefont {Abbott},
  \citenamefont {Abraham}, \citenamefont {Acernese}, \citenamefont {Ackley}
  \emph {et~al.}}]{Abbott_2020}%
  \BibitemOpen
  \bibfield  {author} {\bibinfo {author} {\bibfnamefont {R.}~\bibnamefont
  {Abbott}}, \bibinfo {author} {\bibfnamefont {T.~D.}\ \bibnamefont {Abbott}},
  \bibinfo {author} {\bibfnamefont {S.}~\bibnamefont {Abraham}}, \bibinfo
  {author} {\bibfnamefont {F.}~\bibnamefont {Acernese}}, \bibinfo {author}
  {\bibfnamefont {K.}~\bibnamefont {Ackley}}, \emph {et~al.},\ }\href
  {https://doi.org/10.3847/2041-8213/aba493} {\bibfield  {journal} {\bibinfo
  {journal} {The Astrophysical Journal}\ }\textbf {\bibinfo {volume} {900}},\
  \bibinfo {pages} {L13} (\bibinfo {year} {2020}{\natexlab{b}})}\BibitemShut
  {NoStop}%
\bibitem [{\citenamefont {Field}\ \emph {et~al.}(2014)\citenamefont {Field},
  \citenamefont {Galley}, \citenamefont {Hesthaven}, \citenamefont {Kaye},\
  and\ \citenamefont {Tiglio}}]{PhysRevX.4.031006}%
  \BibitemOpen
  \bibfield  {author} {\bibinfo {author} {\bibfnamefont {S.~E.}\ \bibnamefont
  {Field}}, \bibinfo {author} {\bibfnamefont {C.~R.}\ \bibnamefont {Galley}},
  \bibinfo {author} {\bibfnamefont {J.~S.}\ \bibnamefont {Hesthaven}}, \bibinfo
  {author} {\bibfnamefont {J.}~\bibnamefont {Kaye}},\ and\ \bibinfo {author}
  {\bibfnamefont {M.}~\bibnamefont {Tiglio}},\ }\href
  {https://doi.org/10.1103/PhysRevX.4.031006} {\bibfield  {journal} {\bibinfo
  {journal} {Phys. Rev. X}\ }\textbf {\bibinfo {volume} {4}},\ \bibinfo {pages}
  {031006} (\bibinfo {year} {2014})}\BibitemShut {NoStop}%
\bibitem [{\citenamefont {{Varma}}\ \emph {et~al.}(2019)\citenamefont
  {{Varma}}, \citenamefont {{Field}}, \citenamefont {{Scheel}}, \citenamefont
  {{Blackman}}, \citenamefont {{Gerosa}}, \citenamefont {{Stein}},
  \citenamefont {{Kidder}},\ and\ \citenamefont
  {{Pfeiffer}}}]{2019PhRvR...1c3015V}%
  \BibitemOpen
  \bibfield  {author} {\bibinfo {author} {\bibfnamefont {V.}~\bibnamefont
  {{Varma}}}, \bibinfo {author} {\bibfnamefont {S.~E.}\ \bibnamefont
  {{Field}}}, \bibinfo {author} {\bibfnamefont {M.~A.}\ \bibnamefont
  {{Scheel}}}, \bibinfo {author} {\bibfnamefont {J.}~\bibnamefont
  {{Blackman}}}, \bibinfo {author} {\bibfnamefont {D.}~\bibnamefont
  {{Gerosa}}}, \bibinfo {author} {\bibfnamefont {L.~C.}\ \bibnamefont
  {{Stein}}}, \bibinfo {author} {\bibfnamefont {L.~E.}\ \bibnamefont
  {{Kidder}}},\ and\ \bibinfo {author} {\bibfnamefont {H.~P.}\ \bibnamefont
  {{Pfeiffer}}},\ }\href {https://doi.org/10.1103/PhysRevResearch.1.033015}
  {\bibfield  {journal} {\bibinfo  {journal} {Physical Review Research}\
  }\textbf {\bibinfo {volume} {1}},\ \bibinfo {eid} {033015} (\bibinfo {year}
  {2019})},\ \Eprint {https://arxiv.org/abs/1905.09300} {arXiv:1905.09300
  [gr-qc]} \BibitemShut {NoStop}%
\bibitem [{\citenamefont {Kechedzhi}\ \emph {et~al.}(2023)\citenamefont
  {Kechedzhi}, \citenamefont {Isakov}, \citenamefont {Mandrà}, \citenamefont
  {Villalonga}, \citenamefont {Mi}, \citenamefont {Boixo},\ and\ \citenamefont
  {Smelyanskiy}}]{kechedzhi2023effective}%
  \BibitemOpen
  \bibfield  {author} {\bibinfo {author} {\bibfnamefont {K.}~\bibnamefont
  {Kechedzhi}}, \bibinfo {author} {\bibfnamefont {S.~V.}\ \bibnamefont
  {Isakov}}, \bibinfo {author} {\bibfnamefont {S.}~\bibnamefont {Mandrà}},
  \bibinfo {author} {\bibfnamefont {B.}~\bibnamefont {Villalonga}}, \bibinfo
  {author} {\bibfnamefont {X.}~\bibnamefont {Mi}}, \bibinfo {author}
  {\bibfnamefont {S.}~\bibnamefont {Boixo}},\ and\ \bibinfo {author}
  {\bibfnamefont {V.}~\bibnamefont {Smelyanskiy}},\ }\href@noop {} {\bibinfo
  {title} {Effective quantum volume, fidelity and computational cost of noisy
  quantum processing experiments}} (\bibinfo {year} {2023}),\ \Eprint
  {https://arxiv.org/abs/2306.15970} {arXiv:2306.15970 [quant-ph]} \BibitemShut
  {NoStop}%
\bibitem [{\citenamefont {Abbott}\ \emph
  {et~al.}(2021{\natexlab{b}})\citenamefont {Abbott}, \citenamefont {Abbott},
  \citenamefont {Abraham}, \citenamefont {Acernese}, \citenamefont {Ackley},
  \citenamefont {Adams}, \citenamefont {Adhikari}, \citenamefont {Adya},
  \citenamefont {Affeldt}, \citenamefont {Agathos}, \citenamefont {Agatsuma},
  \citenamefont {Aggarwal}, \citenamefont {Aguiar} \emph
  {et~al.}}]{RICHABBOTT2021100658}%
  \BibitemOpen
  \bibfield  {author} {\bibinfo {author} {\bibfnamefont {R.}~\bibnamefont
  {Abbott}}, \bibinfo {author} {\bibfnamefont {T.~D.}\ \bibnamefont {Abbott}},
  \bibinfo {author} {\bibfnamefont {S.}~\bibnamefont {Abraham}}, \bibinfo
  {author} {\bibfnamefont {F.}~\bibnamefont {Acernese}}, \bibinfo {author}
  {\bibfnamefont {K.}~\bibnamefont {Ackley}}, \bibinfo {author} {\bibfnamefont
  {C.}~\bibnamefont {Adams}}, \bibinfo {author} {\bibfnamefont {R.~X.}\
  \bibnamefont {Adhikari}}, \bibinfo {author} {\bibfnamefont {V.~B.}\
  \bibnamefont {Adya}}, \bibinfo {author} {\bibfnamefont {C.}~\bibnamefont
  {Affeldt}}, \bibinfo {author} {\bibfnamefont {M.}~\bibnamefont {Agathos}},
  \bibinfo {author} {\bibfnamefont {K.}~\bibnamefont {Agatsuma}}, \bibinfo
  {author} {\bibfnamefont {N.}~\bibnamefont {Aggarwal}}, \bibinfo {author}
  {\bibfnamefont {O.~D.}\ \bibnamefont {Aguiar}}, \emph {et~al.},\ }\href
  {https://doi.org/https://doi.org/10.1016/j.softx.2021.100658} {\bibfield
  {journal} {\bibinfo  {journal} {SoftwareX}\ }\textbf {\bibinfo {volume}
  {13}},\ \bibinfo {pages} {100658} (\bibinfo {year}
  {2021}{\natexlab{b}})}\BibitemShut {NoStop}%
\bibitem [{\citenamefont {Abbott}\ \emph {et~al.}(2023)\citenamefont {Abbott},
  \citenamefont {Abe}, \citenamefont {Acernese}, \citenamefont {Ackley},
  \citenamefont {Adhicary}, \citenamefont {Adhikari}, \citenamefont {Adhikari},
  \citenamefont {Adkins}, \citenamefont {Adya}, \citenamefont {Affeldt} \emph
  {et~al.}}]{Abbott_2023}%
  \BibitemOpen
  \bibfield  {author} {\bibinfo {author} {\bibfnamefont {R.}~\bibnamefont
  {Abbott}}, \bibinfo {author} {\bibfnamefont {H.}~\bibnamefont {Abe}},
  \bibinfo {author} {\bibfnamefont {F.}~\bibnamefont {Acernese}}, \bibinfo
  {author} {\bibfnamefont {K.}~\bibnamefont {Ackley}}, \bibinfo {author}
  {\bibfnamefont {S.}~\bibnamefont {Adhicary}}, \bibinfo {author}
  {\bibfnamefont {N.}~\bibnamefont {Adhikari}}, \bibinfo {author}
  {\bibfnamefont {R.~X.}\ \bibnamefont {Adhikari}}, \bibinfo {author}
  {\bibfnamefont {V.~K.}\ \bibnamefont {Adkins}}, \bibinfo {author}
  {\bibfnamefont {V.~B.}\ \bibnamefont {Adya}}, \bibinfo {author}
  {\bibfnamefont {C.}~\bibnamefont {Affeldt}}, \emph {et~al.},\ }\href
  {https://doi.org/10.3847/1538-4365/acdc9f} {\bibfield  {journal} {\bibinfo
  {journal} {The Astrophysical Journal Supplement Series}\ }\textbf {\bibinfo
  {volume} {267}},\ \bibinfo {pages} {29} (\bibinfo {year} {2023})}\BibitemShut
  {NoStop}%
\bibitem [{\citenamefont {Karnaugh}(1953)}]{6371932}%
  \BibitemOpen
  \bibfield  {author} {\bibinfo {author} {\bibfnamefont {M.}~\bibnamefont
  {Karnaugh}},\ }\href {https://doi.org/10.1109/TCE.1953.6371932} {\bibfield
  {journal} {\bibinfo  {journal} {Transactions of the American Institute of
  Electrical Engineers, Part I: Communication and Electronics}\ }\textbf
  {\bibinfo {volume} {72}},\ \bibinfo {pages} {593} (\bibinfo {year}
  {1953})}\BibitemShut {NoStop}%
\end{thebibliography}%

\newpage

\appendix
\section{The divide and conquer algorithm}
\label{sec:appa}
Here, we describe the divide and conquer data loading algorithm proposed by Araujo et al.~\cite{Araujo_2021}. The algorithm encodes an $n$-dimensional input vector using a tree architecture. The algorithm requires requires $\mathcal{O}(n)$ classical resources and $\mathcal{O}(\log_2^2(n))$ quantum resources to amplitude encode an $n$-dimensional real vector.

The algorithm is based on a bottom-up approach, where qubits are combined in groups at each step. Therefore, the first step of the algorithm requires single qubit states. These states are generated by single qubit $R_Y$ rotations. The computation of the angles of these gates are the source of $\mathcal{O}(n)$ classical resources requirement. Then, a subroutine is required to combine these states. The first level of this step will create two qubit pairs, the second level will combine these to create groups of three qubit states. Therefore, this step requires $\log_2(n)$ levels. 

The subroutine to combine these states will dictate the total depth of the algorithm. Araujo et al.~\cite{Araujo_2021} uses the following circuit to combine two $m$ qubit states.

\begin{figure}[!h]
    \centering
    \begin{quantikz}
        \lstick{a\ket{0}+b\ket{1}} & \ctrl{2}  & \ctrl{2}  & \qw  &\ctrl{2}  & \qw \rstick[7]{a\ket{0}\ket{\psi}\ket{\phi}\\+\\b\ket{1}\ket{\phi}\ket{\psi}}\\   
        \lstick[3]{$\ket{\psi}_m$} & \targX{}  & \qw        & \midstick[3,brackets=none]{\vdots} &\qw       & \qw \\       
        & \qw       & \targX{}  &   & \qw       & \qw \\[0.3cm]     
        & \qw       & \qw       &   & \targX{}  & \qw \\             
        \lstick[3]{$\ket{\phi}_m$} & \swap{-3} & \qw        & \midstick[3,brackets=none]{\vdots} & \qw       & \qw \\      
        & \qw       & \swap{-4} &   & \qw       & \qw \\[0.3cm]      
        & \qw       & \qw        &  & \swap{-4} & \qw          
    \end{quantikz}
    \caption{Circuit to combine two $m$-qubit states. a and b do not depend on the input states $\ket{\psi}$ and $\ket{\phi}$. The depth of the subroutine is $\mathcal{O}(m)$ and $m$ qubits are discarded at each step.}
    \label{fig:my_label}
\end{figure}
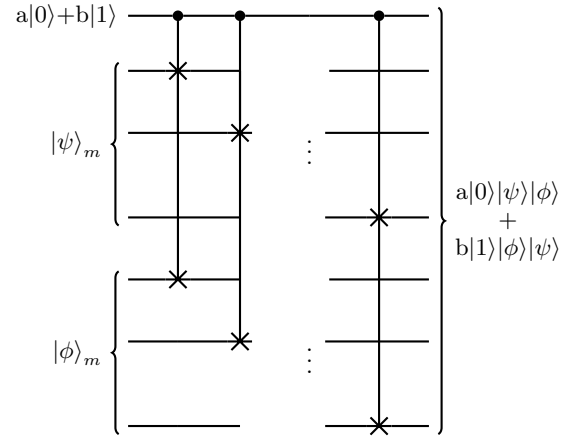

This subroutine is used to combine the states sequentially, such that first level combines one qubit states, $m^{\rm th}$ level combines m qubit states and the last level combines $\log_2(n)-1$ qubit states to encode the $n$-dimensional real vector. Thus, the depth of the algorithm is $\mathcal{O}(\log_2^2(n))$ and it uses $O(n)$ qubits in total including the auxiliary qubits. Please refer to Araujo et al.~\cite{Araujo_2021} for more details.

\section{Time complexity and resource analysis}
\label{sec:appb}
Our method is based on encoding data and signal templates, measuring the qubits, doing classical logic operations on the measurements and doing mathematical corrections at the end. For obtaining a fixed precision, excluding the pre and post processing parts, the quantum and classical logic circuits need to run multiple times. This repeat factor is proportional to the length of the data $\mathcal{O}(L)$ since every quantum measurement outcome is distributed to one of the $L$ possible times. Time complexity of each stage is:
\begin{itemize}
    \item \textbf{Pre-encoding calculations:} Encoding requires calculating the state amplitudes to be encoded. This corresponds to calculating the offsets $\Delta x$ and $\Delta y$, taking the square root and calculating the normalizations. The time complexity of these operations are proportional to the length of the data, i.e. $\mathcal{O}(L)$. This needs to be done once.

Second part of the pre-encoding stage is the design of the encoding circuit. This can be done with time complexity $\mathcal{O}(L)$ for the divide and conquer algorithm we use for encoding~\cite{Araujo_2021}. This needs to be done once.
 \item \textbf{Encoding (quantum):} 
With the divide and conquer algorithm this is done in $\mathcal{O}((\log L)^2)$. This is done $\mathcal{O}(L)$ times. 
\item \textbf{Constructing and adding products (classical):}There are going to be total of $\log_2NL$ qubits to be measured and $NL$ possible different outcomes. SNR of each data point $j$ is proportional to the number of measurements of mutually exclusive $N$ of these possible different outcomes. Therefore depth of the logic circuits constructed with OR gates with 2 inputs are $\mathcal{O}(\log_2L)$. 

The expansion of $\log_2NL$ qubits to $NL$ different logic functions can be done with 2 input AND gates and inverters with depth $\mathcal{O}(\log\log NL)$. These two operations in series will be done $\mathcal{O}(L)$ times.

\item \textbf{Post-measurement calculations: } Among the correction terms in Eq. \eqref{eq:corrected} only the $\sum_{i=1}^N\Delta xy_{i+j}$ term is not a constant term. Although it seems that $N$ additions need to be made for calculating it for every $L$ values of $j$, since the difference between the consecutive values of the correction is just 2 numbers (first number of the previous step and the last number of the new step), asymptotically $\mathcal{O}(N+L)$ operations are needed for calculating every $L$ value of it.
\end{itemize}

Overall time complexity is dominated by the encoding part, which is $\mathcal{O}(L(\log L)^2)$

\textbf{Extra optimization: }Since the overall scaling $\mathcal{O}(L(\log L)^2)$ grows faster than linear scaling $\mathcal{O}(L)$, one might suggest to divide the data into smaller segments with length $k$ and perform several smaller computations in series. In order to have completeness of SNRs, each of these segments must overlap with the previous one at the neighboring $N$ data points. Then the computation time becomes $\mathcal{O}(\frac{k}{k-N}L(\log k)^2)$. The optimal length $k$ depends on the template length which is given by the equation $N=\frac{2k}{\ln k+2}$. This optimization scheme decreases the total computation time generally to $\mathcal{O}(L(\log N)^2)$. The precise optimization does not affect the asymptotic scaling. For example, with the choices of $k=2N$ or $k=3N$, it becomes $\mathcal{O}(2L(\log 2N)^2)$ or $\mathcal{O}(1.5L(\log 3N)^2)$ respectively, both being equivalent to $\mathcal{O}(L(\log N)^2)$.

\textbf{Hardware requirements: }Total necessary qubit count is $\mathcal{O}(L)$. Total number of needed AND gates and inverters, without any logical simplifications, e.g., with a Karnaugh map~\cite{6371932}; will be $\mathcal{O}(NL\log NL)$. Similarly, total needed OR gates will be $\mathcal{O}(NL)$.

\textbf{Precision of the computation: } Without the corrections in Eq. \eqref{eq:corrected}, according to variance and mean of the binomial distribution, the relative precision of the SNR ($\rho[j]$) estimates would be (standard deviation/mean)
\begin{equation}
\label{eq:ver}
 \frac{\delta\rho[j]}{\rho[j]}=\sqrt{\frac{\sum_{i\neq j}\rho[i]}{\rho[j]\times s\times L}}\approx \sqrt{\frac{\Bar{\rho}}{\rho[j]\times s}}
 \end{equation}
 where $s$ is the total shot count divided by $L$, which is taken as constant in the above time complexity analysis; and $\Bar{\rho}$ is the average SNR. With the corrections, this relative precision becomes the precision of SNR$+\sum_{i=1}^N\Delta yx_i+\Delta xy_{i+j}+\Delta y\Delta x$. Therefore the relative precision of the SNRs become
\begin{multline}
    \frac{\delta \rho[j]}{\rho[j]}=\sqrt{\frac{\sum_{i\neq j}\rho[i]}{\rho[j]\times s\times L}}\\ \times\left(1+\frac{\sum_{i=1}^N\Delta yx_i+\Delta xy_{i+j}+\Delta y\Delta x}{\rho[j]}\right)
\end{multline}
The offsets worsen the precision, and each data segment with different offsets can have different precision with the same shot count.

\section{Verification with artificial random data}
\label{sec:appc}
\begin{figure*}
    \centering
    \includegraphics[width=\textwidth]{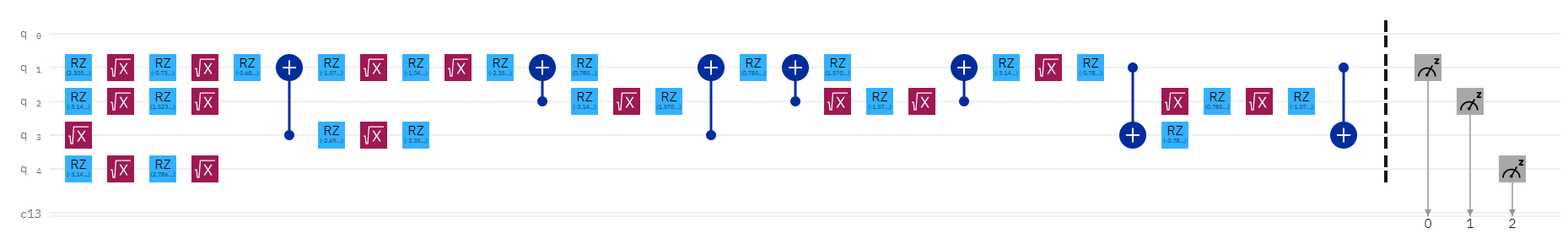}
    \caption{One of the circuits ran in the experiments for encoding}
    \label{fig:circ}
\end{figure*}

Prior to anlayzing real data, we tested the accuracy of our method on artificially produced random data. The experiments we show here were ran on the backend {\it ibmq\_lima}. The estimated total CNOT error probability and the total measurement error probabilities were both about 7\% which are the main estimated sources of error. We have chosen the 2 point signal template arbitrarily as [2,-1]. Each data set of 4 numbers were chosen randomly from a normal distribution of zero mean and unit variance. In order to have all of the possible 8 measurement outcomes, the signal and data were arbitrarily shifted  0.1 more than their minimum negative values, i.e. $\Delta x=-{\rm min}(x_i)+0.1$, $\Delta y=-{\rm min}(y_i)+0.1$. Here we show results from 100 different data sets for each 2$\times10^4$ measurements (shots) were made. Fig.~\ref{fig:circ} shows an example circuit from our experiment for the encoding of 3 qubits and their measurements. Qubits q1, q2, and q3 are used for encoding the data into the qubits q1 and q2. The signal template is encoded to q4.

\begin{figure*}
    \centering
    \includegraphics[width=\textwidth]{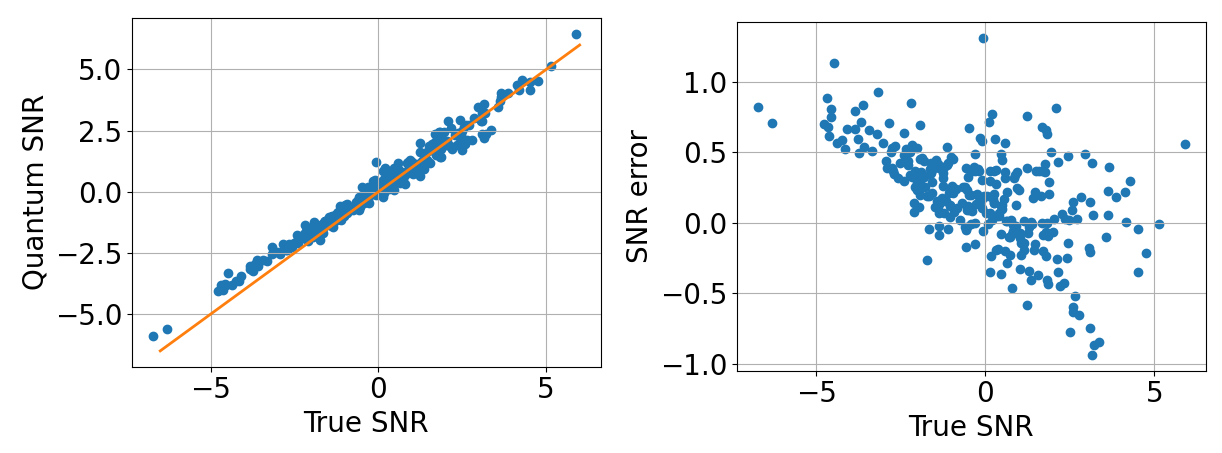}
    \caption{Results of the experiments on the quantum backend {\it ibmq\_lima}: Left figure shows a scatter of the 300 SNRs computed with the use of qubits vs the true SNRs. The orange line is x=y just for reference. Right figure shows the dependency on the errors in the computation on the true SNRs.}
    \label{fig:lima}
\end{figure*}

\begin{figure*}
    \centering
    \includegraphics[width=\textwidth]{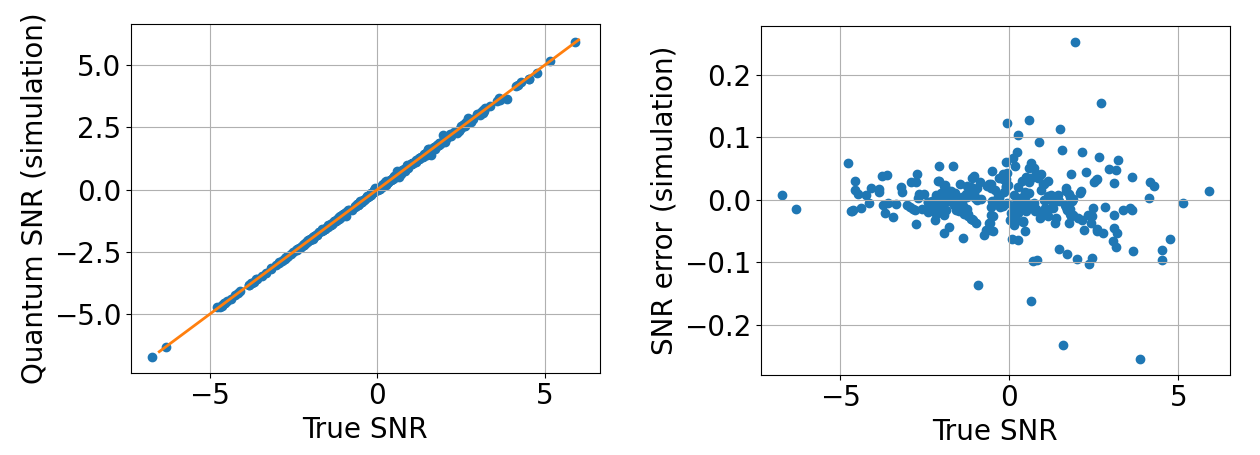}
    \caption{Results from an ideal noise free quantum simulation. The orange line is again x=y just for reference. The only source of error is the Poisson uncertainty due to finite number of shots=$2\times10^4$.}
    \label{fig:sim}
\end{figure*}

Fig.~\ref{fig:lima} shows two scatter plots, one for the calculated $100\times3=300$ SNRs with quantum measurements vs. the true SNR values and one for the SNR errors between them vs. the true SNRs. The correlation coefficient for the points in the left figure was found to be 0.99, which should have been 1 ideally, and in the right figure as -0.57. The anti-correlation between the errors and the true SNR is a clear indication that the errors have arisen due to the noise in the circuit. This is due to the fact that in order to have an SNR with a high magnitude, either the amplitude of the data should be high or more relevantly the relative amplitudes of the consequent data points need to have specific values. Any noise in the system can corrupt such delicate data segments. These corruptions increase the marginal entropies of encoded signal and data incoherently resulting in decrease in their mutual information. Low SNR points on the contrary do not get affected by such corruptions as their already low mutual information cannot decrease more. Another observation that can be made is the affinity to having positive SNR errors. This can be explained by the asymmetry in the encoding due to shifting to the positive values. Without the correction in Eq.~\eqref{eq:corrected}, only the positive SNR values can be obtained. Therefore there is a fundamental lower limit on the SNR errors during the encoding. The effect of this lower limit is seen as having mostly positive SNR errors, especially for data segments which have true negative SNR since their SNR values without the correction are the closest to zero. In Fig.~\ref{fig:sim} we show the result of a noise-free ideal quantum simulation with the same data and signal. The error in this case is only due to the Poisson errors due to finite number of measurements. The correlation coefficient between the SNRs is 1.0 and between the errors and the true SNRs is -0.1.

\end{document}